\documentclass[aps,prb,twocolumn,amsmath,superscriptaddress,footinbib]{revtex4}
\usepackage{graphicx}
\usepackage{color}
\usepackage{epstopdf}

\usepackage{xcolor}

\begin{document}

\title{A hole-Cr$^{+}$ nano-magnet in a semiconductor quantum dot}

\author{V.~Tiwari}
\affiliation{Institut N\'{e}el, CNRS, Univ. Grenoble Alpes and Grenoble INP, 38000 Grenoble, France}

\author{M.~Arino}
\affiliation{Institute of Materials Science, University of Tsukuba, Tsukuba 305-8573, Japan}

\author{S.~Gupta}
\affiliation{Department of Physics, IriG, Univ. Grenoble Alpes and CEA, 38000 Grenoble, France}

\author{M.~Morita}
\affiliation{Institute of Materials Science, University of Tsukuba, Tsukuba 305-8573, Japan}

\author{T.~Inoue}
\affiliation{Institute of Materials Science, University of Tsukuba, Tsukuba 305-8573, Japan}

\author{D.~Caliste}
\affiliation{Department of Physics, IriG, Univ. Grenoble Alpes and CEA, 38000 Grenoble, France}

\author{P.~Pochet}
\affiliation{Department of Physics, IriG, Univ. Grenoble Alpes and CEA, 38000 Grenoble, France}

\author{H.~Boukari}
\affiliation{Institut N\'{e}el, CNRS, Univ. Grenoble Alpes and Grenoble INP, 38000 Grenoble, France}

\author{S.~Kuroda}
\affiliation{Institute of Materials Science, University of Tsukuba, Tsukuba 305-8573, Japan}

\author{L.~Besombes}\email{lucien.besombes@neel.cnrs.fr}
\affiliation{Institut N\'{e}el, CNRS, Univ. Grenoble Alpes and Grenoble INP, 38000 Grenoble, France}


\begin{abstract}

We study a new diluted magnetic semiconductor system based on the spin of the ionized acceptor Cr$^+$. We show that the negatively charged Cr$^+$ ion, an excited state of the Cr in II-VI semiconductor, can be stable when inserted in a CdTe quantum dot (QD). The Cr$^+$ attracts a heavy-hole in the QD and form a stable hole-Cr$^+$ complex. Optical probing of this system reveals a ferromagnetic coupling between heavy-holes and Cr$^+$ spins. At low temperature, the thermalization on the ground state of the hole-Cr$^+$ system with parallel spins prevents the optical recombination of the excess electron on the 3$d$ shell of the atom. We study the dynamics of the nano-magnet formed by the hole-Cr$^+$ exchange interaction. The ferromagnetic ground states with M$_z$=$\pm$4 can be controlled by resonant optical pumping and a spin relaxation time in the 20 $\mu$s range is obtained at T=4.2 K. This spin memory at zero magnetic field is limited by the interaction with phonons. 

\end{abstract}

\maketitle

Recent experimental breakthroughs have laid the foundations for atomic scale data storage, showing the capability to read and control the spin of a single magnetic atom. In particular, a magnetic atom can develop a magnetic anisotropy energy when it interacts with the surface of a metal and behave like a nano-magnet \cite{Oberg2014}. A single magnetic atom can also present a large magnetic anisotropy and a spin memory when interacting with ligands in a molecular magnet \cite{Wernsdorfer2008}. In semiconductors, the optical properties of a quantum dot (QD) can be used to control the spin of individual magnetic atoms \cite{Besombes2004,Kudelski2007,Krebs2009,Kobak2014,Besombes2012,Krebs2013,Varghese2014,Bacher2016}. Embedding a magnetic atom in a QD offers in addition the possibility to tune the environment of the localized spin. A control of the charge of the QD can for instance influence the magnetic anisotropy of the atom \cite{Lafuente2017}, making these nano-sized systems attractive for miniaturized data storage applications.

A variety of magnetic transitions metals can be incorporated in semiconductors offering a large choice of localized electronic spins, nuclear spins as well as orbital momentum. The 3$d^5$ Mn$^{2+}$, a pure spin without orbital momentum, is the most widely studied magnetic element in nano-structures \cite{Bacher2020,Bayer2020}. The magnetic properties of these nano-structures are mainly controlled by the exchange interaction of the holes spins with the 5 $d$ electrons of the atom. In a II-VI compounds, this exchange interaction is dominated by $p-d$ hybridization, the so-called kinetic exchange with anti-ferromagnetic sign \cite{Furdyna1988,Kacman2001,Kossut,Beaulac2010}. This anti-ferromagnetic interaction limits the spin stability of the complex formed by a heavy hole and one or a few Mn atoms as hole-Mn spins flip-flops occur in the magnetic ground state \cite{Lafuente2017}.

We demonstrate here that another 3$d^5$ element, the  negativelly charged ionized acceptor Cr$^+$, can be optically probed when inserted in CdTe/ZnTe QDs. We study the spin properties of this new diluted magnetic semiconductor system. The negative charge of the Cr$^+$ ion attracts a hole in the QD and the hole-Cr$^+$ complex is formed. Magneto-optics measurements on these positively charged QDs show that the hole-Cr$^+$ exchange interaction is ferromagnetic. This coupling leads to a ground state configuration with parallel hole and Cr$^+$ spins which blocks the recombination of the excess electron of the 3$d$ shell. The two low energy hole-Cr$^+$ states, with angular momentum M$_z$ = $\pm$4, behave like an Ising spin system that can thus be seen as a nano-magnet. Resonant optical pumping is used to probe the dynamics of the hole-Cr$^+$ complex. A spin memory in the 20 $\mu s$ range is observed at T=4.2 K and zero magnetic field. This dynamics is controlled by the interaction with phonons.

For these experiments, Cr atoms are randomly introduced in CdTe/ZnTe QDs grown by molecular beam epitaxy on a 1 $\mu m$ thick ZnTe buffer layer deposited on a GaAs (001) substrate \cite{Wojnar2011}. The amount of Cr is adjusted to obtain QDs containing 0, 1 or a few Cr atoms. Individual QDs are studied in magnetic field by optical micro-spectroscopy \cite{Supplemental}.

\begin{figure}[hbt]
\centering
\includegraphics[width=1.0\linewidth]{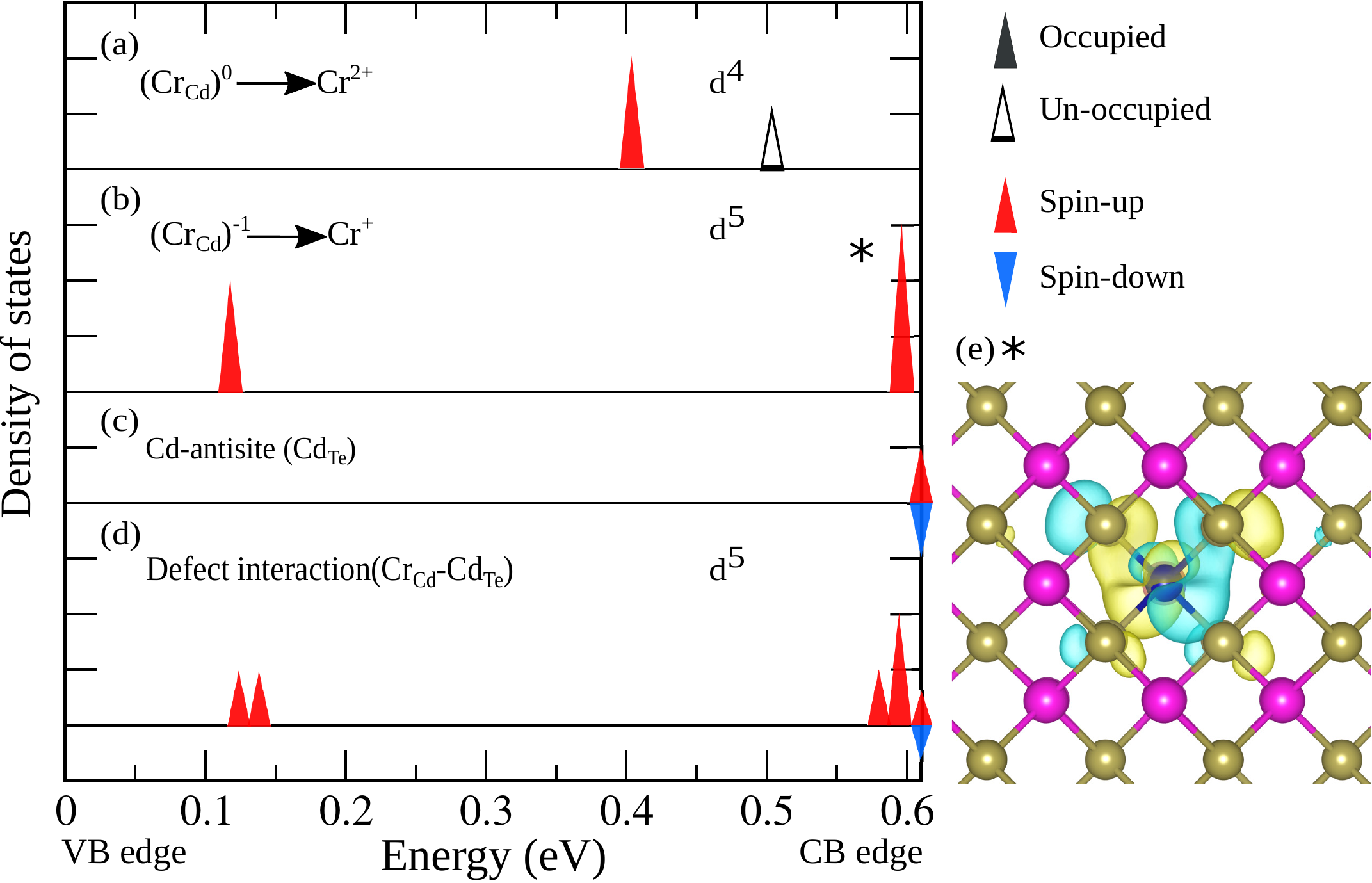}
\caption{DFT calculated defects electronic levels in the band gap of CdTe for a Cr$^{2+}$ (a), a Cr$^+$ (b) along with (c) a Cd antisite compared to a configuration (d) where the Cr$_{Cd}$ and Cd$_{Te}$ are separated by 9.5 $\AA$. (e) Calculated orbital of one of the state in the high energy triplet of the Cr$^+$ showing the hybridization of Cr and Te orbitals.}
\label{Fig1bis}
\end{figure}

A free Cr atom exhibits a 3$d^5$4$s^1$ electron configuration. In the cation site of a II-VI semiconductor, two of its electrons are given to the bonds. The iso-electronic impurity Cr$^{2+}$ (3$d^4$) is formed with a spin S=2 and an orbital momentum L=2. However, in CdTe, Cr$^{2+}$ is also an acceptor \cite{Cieplak1975,Godlewski1980,Besombes2019fluc}. This is illustrated in Fig.~\ref{Fig1bis} presenting Density Functional Theory based calculations of substitutional Cr in CdTe (DFT code BigDFT \cite{Supplemental,Genovese2008,Mohr2015,Hartwigsen1998}). A Cr sitting on a Cd-site is adopting a 2+ oxidation state with two occupied levels in the valence band, two occupied and one unoccupied levels below the conduction band edge (Fig.~\ref{Fig1bis}(a)). The Cr can then capture an electron on its unoccupied level; it adopts a 1+ oxidation state (3$d^5$, S=5/2 and L=0) with five localized levels in the band gap, a low energy doublet and a hight energy triplet (Fig.~\ref{Fig1bis}(b)). The triplets states are significantly mixed with the anion $p$-orbitals (Fig.~\ref{Fig1bis}(e)). This $p-d$ hybridization is a potential source of kinetic exchange between the Cr$^+$ and holes spins \cite{Furdyna1988,Kacman2001,Kossut,Beaulac2010}.

The negatively charged acceptor Cr$^+$ is however unstable in bulk II-VI materials. Its magnetic properties and exchange interaction with the carriers of the host have never been studied. The presence of Cr$^+$ was nevertheless detected in electron paramagnetic resonance experiments in bulk n-type CdTe or under optical excitation \cite{Cieplak1975,Godlewski1980}. 

We used DFT calculations to evaluate the influence of defects on the possibility to observe Cr$^+$ in undoped CdTe. Indeed, charge compensation with intrinsic point defects is a common mechanism \cite{Decapito2013}. We found that a Cd anti-site (donor character) can give rise to a change in the oxidation state of the Cr. Fig.~\ref{Fig1bis} depicts the defects states for isolated Cr and Cd anti-site as compared to a configuration where they are located at a distance of 9.5 $\AA$ (3$^{rd}$ neighbor). The latter configuration presents five states in the band gap with five unpaired electrons (Fig.~\ref{Fig1bis}(d)). Even at this small distance, the electronic states are located on the Cr center with energies and orbitals very similar to a 3$d^5$ isolated Cr$^+$. Similarly, Zn anti-site in the barriers of Cr-doped CdTe/ZnTe QDs could be a local source of electron transfer to Cr$^{2+}$ by modulation doping.

QDs containing an individual iso-electronic Cr$^{2+}$ were identified recently \cite{Lafuente2018}. In these dots, the spin S=2 of Cr$^{2+}$ is split by a large fine structure term induced by local strain. At low temperature, the exchange interaction with a confined exciton gives rise to three main emission lines corresponding to the lowest energy spin states of the atom S$_z$=0 and S$_z$=$\pm1$. 

\begin{figure}[hbt]
\centering
\includegraphics[width=1.0\linewidth]{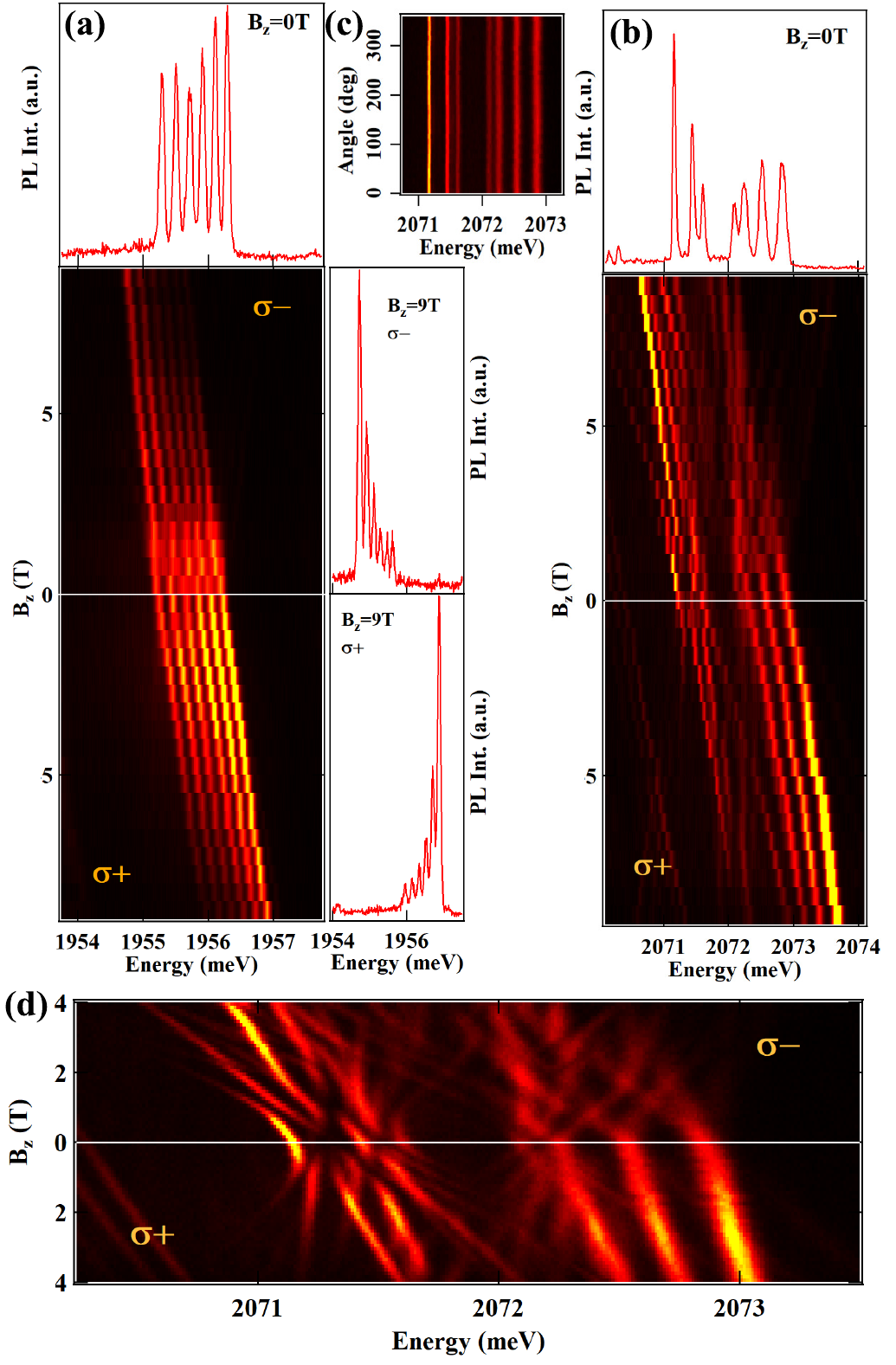}
\caption{Magnetic field dependence of the PL of two positively charged QDs, QD1 (a) and QD2 (b) containing a Cr$^+$. The bottom panels present PL intensity maps recorded in circular polarization. (c) Linear polarization PL intensity map of QD2 at B$_z$=0T. (d) Detail of the PL intensity map of QD2.}
\label{Fig1}
\end{figure}

Fig.~\ref{Fig1} presents the magnetic field dependence of another type of QDs observed in Cr-doped samples. Their emission consists in a minimum of six lines equally spaced in energy (see QD1)\cite{note1}. Most of these QDs present a more complex structure with the third line clearly split and seven lines separated by a gap (see QD2). The width of the gap changes from dot to dot and additional weak intensity lines can also be observed \cite{Supplemental}. No linear polarisation was observed in any of the investigated QDs (Fig.~\ref{Fig1}(c)) suggesting that this PL structure arises from a charged exciton ({\it i.e.} absence of fine structure splitting) \cite{Leger2007}.

Under magnetic field each line presents a Zeeman splitting and a change in the photoluminescence (PL) intensity distribution is observed. At B$_z$=9T the PL is concentrated on the low energy lines in $\sigma-$ polarization and on the high energy lines in $\sigma+$ polarization (Fig.~\ref{Fig1}(a)). This intensity distribution is influenced by the temperature and the excitation power and is opposite to the situation reported for Mn$^{2+}$ in II-VI QDs \cite{Besombes2004,Supplemental}. A line broadening or anti-crossings are usually observed in the low magnetic field region. This is particularly pronounced for QD2 where a complex series of anti-crossings is observed below B$_z$=4T (Fig.~\ref{Fig1}(d)).

This PL structure can be explained by the recombination of a positively charged exciton (X$^+$) interacting with the spin S=5/2 of the Cr in its 3$d^5$ configuration, the Cr$^+$ ionized acceptor. This charged ion, when located in a QD, attracts a hole and repels electrons. Thus, a hole-Cr$^+$ complex is formed that can be probed by the optical injection of an exciton.

\begin{figure}[hbt]
\centering
\includegraphics[width=1.0\linewidth]{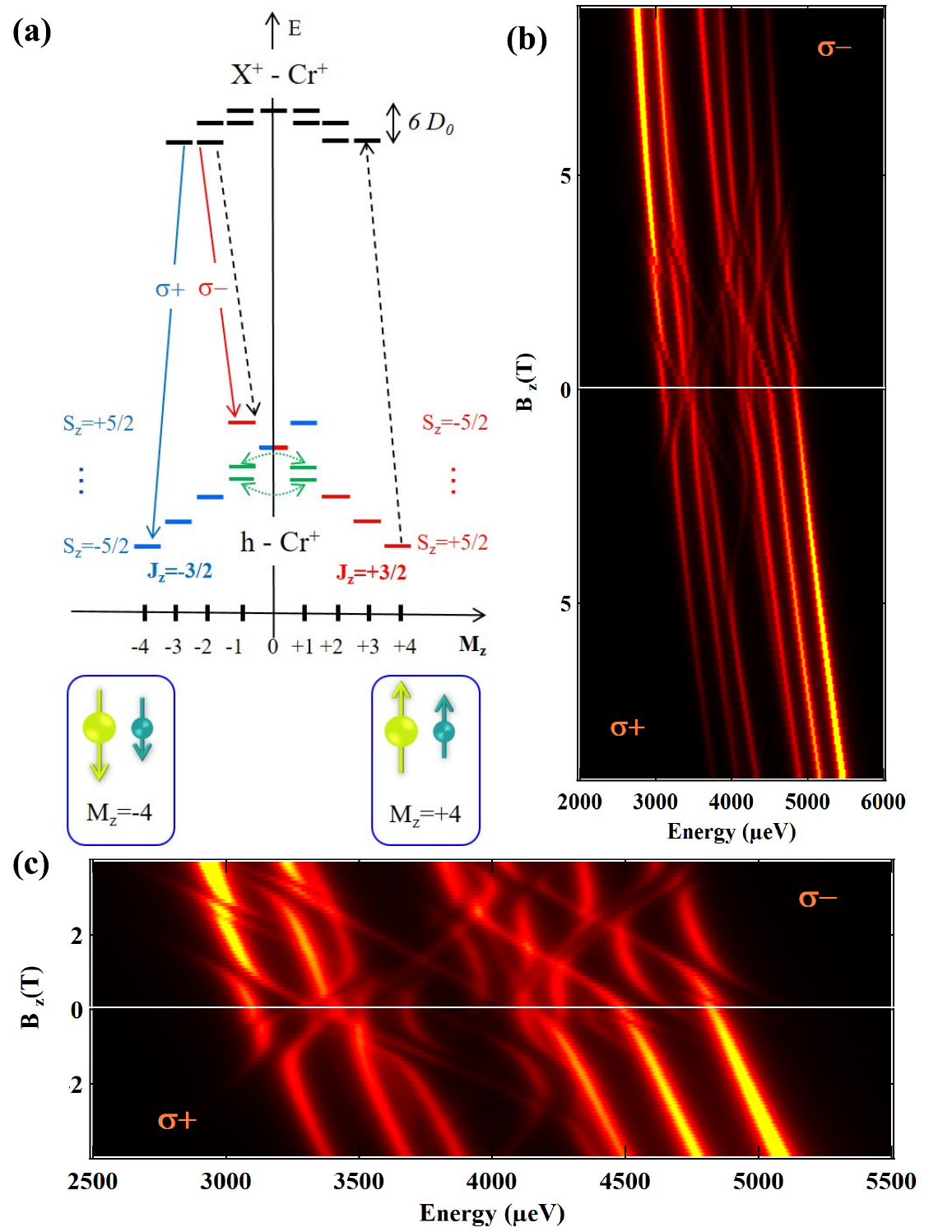}
\caption{(a) Energy levels of the excited state (X$^+$-Cr$^+$) and ground state (h-Cr$^+$) in a positively charged Cr$^+$-doped QD. The electron-Cr exchange interaction and the fine structure term $E$ are neglected and the structure of X$^+$-Cr$^+$ is controlled by D$_0$. The most intense transitions in $\sigma+$ and $\sigma-$ polarizations are indicated with thick arrows. Dashed arrows show the excitation/detection configuration for the pumping experiment. (b) PL intensity map of $X^+-Cr^+$ calculated with $I_{eCr^+}=0\mu eV$, $I_{hCr^+}=-225\mu eV$, $D_0=-40 \mu eV$, $E=20 \mu eV$, $T_{eff}=20K$, $\rho/\Delta_{lh}=0.2$, $\theta_s=-\pi/4$ and $\gamma$=1.5 $\mu eV T^{-2}$, g$_{Cr^+}$=2, g$_{h}$=0.5, g$_{e}$=-0.4. (c) Detail of the calculated PL intensity map.}
\label{Fig2}
\end{figure}

This attribution is confirmed by the modelling of the magnetic field dependence of X$^+$-Cr$^+$ presented in Fig. \ref{Fig2}. In this model, we consider that the two holes are not localized on the negatively charged acceptor and keep a heavy-hove character with a spin J$_z$=$\pm$3/2 \cite{NoteLocal}. The energy levels of X$^+$-Cr$^+$ are then described by the Hamiltonian

\begin{eqnarray}
{\cal H}_{X^+-Cr^+}=I_{eCr^+}\vec{S}\cdot\vec{\sigma}+g_{e}\mu_B\vec{\sigma}\cdot\vec{B}+\gamma B^2+{\cal H}_{Cr^+}
\end{eqnarray}

\noindent which contains the exchange interaction between the Cr spin ($\vec{S}$) and the electron spin ($\vec{\sigma}$), the Zeeman energy of the electron spin and a quadratic diamagnetic shift. ${\cal H}_{Cr^+}=D_0S^2_z+E(S_x^2-S_y^2)+g_{Cr^+}\mu_B\vec{S}\cdot\vec{B}$ contains the fine structure of the Cr$^+$ spin and its Zeeman energy with $g_{Cr^+}\approx2$ \cite{Godlewski1980,Ludwig1963}. The strain induced fine structure consists of a biaxial term $D_0$ and a term arising from a possible in-plane anisotropy $E$. The exchange interaction of the two spin paired holes with the Cr$^+$ is neglected \cite{Hawrylak2013,Lafuente2015}.

For $I_{eCr} \gg D_0$ and $E$, ${\cal H}_{X^+-Cr^+}$ describes the isotropic coupling of a spin 5/2 and a spin 1/2. This would result in two energy levels with a total angular momentum M=2 or M=3 and split by 3I$_{eCr}$. For a vanishingly small I$_{eCr}$, energies of X$^+$-Cr$^+$ are dominated by the fine structure ${\cal H}_{Cr^+}$.

The hole-Cr$^+$ complex in the ground state is described by 

\begin{eqnarray}
{\cal H}_{h-Cr^+}=I_{hCr^+}\vec{S}\cdot\vec{J}+g_{h}\mu_B\vec{J}\cdot\vec{B}+{\cal H}_{Cr^+}
\end{eqnarray}

\noindent where $\vec{J}$ is the hole spin operator. In the subspace of the two low-energy heavy-hole states, a pseudo-spin operator $\vec{\tilde{j}}$ can be used to take into account a possible influence of the valence band mixing (VBM). For a VBM induced by in-plane anisotropy of the strain, the $\vec{\tilde{j}}$ components are related to the Pauli matrices $\tau$ by $\tilde{j}_z=\frac{3}{2}\tau_z$ and $\tilde{j}_{\pm}=\xi\tau_{\pm}$ with $\xi=-2\sqrt{3}\rho_s/\Delta_{lh}\exp{(-2i\theta_s)}$. $\rho_s$ is the coupling energy between heavy-holes and light-holes split by the energy $\Delta_{lh}$, and $\theta_s$ is the angle relative to the (100) axis describing the anisotropy responsible for the VBM \cite{Supplemental}.

Neglecting the VBM ($\rho_s/\Delta_{lh}\approx 0$), the 12 eigenstates of ${\cal H}_{h-Cr^+}$ are organized as six equally spaced doublets with defined $S_z$ and $J_z$. For each level of X$^+$-Cr$^+$ with either M=2 or M=3 there are six possible final states after annihilation of an electron-hole pair. With 3I$_{eCr}$ larger than the width of the emission lines (around 75 $\mu$eV) we would expect 12 spectrally resolved PL lines for X$^+$-Cr$^+$. For a low value of I$_{eCr}$ (lower than a few $\mu eV$) and with $E \ll D_0$, S$_z$ is a good quantum number in the excited state ({\it i.e.} electron-Cr$^+$ states), it is conserved during the optical transition and only six lines are obtained, as observed for QD1.

For most of the QDs, the third line is split and an energy gap is observed in the center of the X$^+$-Cr$^+$ PL spectra (QD2 in Fig.~\ref{Fig1}) \cite{Supplemental}. This structure results from (i) the presence of a VBM induced by an in-plane anisotropy which couples two by two the hole-Cr$^+$ levels and (ii) a weak electron-Cr$^+$ exchange interaction. 

Provided that $\rho_s / \Delta_{lh} \ll 1$, the effect of the VBM is small on the degeneracy of all the hole-Cr$^+$ doublets except for the fourth which is split (see Fig \ref{Fig2}(a)). The split states are the bonding and anti-bonding combinations of $\vert S_z=-1/2,J_z=+3/2\rangle$ and $\vert S_z=+1/2,J_z=-3/2\rangle$. For 3I$_{eCr}$ larger than the width of the PL lines the optical transition to these mixed states would give rise to linearly polarized lines \cite{Varghese2014}. For a low value of I$_{eCr}$, the two linearly polarized transitions are degenerated, only seven lines are expected (see QD2) and the width of the central gap is controlled by $\rho_s/\Delta_{lh}$. 

The distribution of PL intensity under magnetic field is opposite to the situation of a Mn$^{2+}$ ion in similar CdTe/ZnTe QDs. In these dots, the level structure is dominated by an anti-ferromagnetic hole-Mn$^{2+}$ exchange interaction \cite{Besombes2004}. To reproduce the magnetic field dependence of the intensity distribution of X$^+$-Cr$^+$ a ferromagnetic hole-Cr$^+$ exchange interaction and a thermalisation on the electron-Cr$^+$ states ({\it i.e.} initial states) with an effective temperature $T_{eff}$ are used. Based on the perturbation approach presented in ref. 15, a ferromagnetic coupling is indeed expected for a 3d$^5$ element with the hybridized triplet states (star in Fig.\ref{Fig1bis}) situated above the edge of the valence band (see Supp. Mat. \cite{Supplemental}). A value of I$_{eCr}$ lower than a few $\mu$eV is also required to obtain the structure of six (or seven) lines. For $\vert I_{hCr^+}\vert\gg\vert I_{eCr}\vert\approx0$, the hole-Cr$^+$ exchange interaction can be directly deduced from the overall splitting of the emission spectra given by 3/2$\times 5I_{hCr^+}$. 

For I$_{eCr}$ lower than $D_0$ and $E$, the energy levels of the X$^+$-Cr$^+$ are controlled by ${\cal H}_{Cr^+}$. In the weak magnetic field region where $g_{Cr^+}\mu_B B \leq 6 D_0$, overlaps of the different X$^+$-Cr$^+$ spin levels are possible. These states can be mixed by the $E$ term producing anti-crossings in the initial state of the optical transitions. A value of E$\approx 20 \mu eV$ is required to reproduce the spectra of QD2 below B$_z$=4T (Fig.~\ref{Fig1}(d) and Fig.~\ref{Fig2}(c)). D$_0$ shifts the magnetic field position of these anti-crossings. A negative D$_0$ in a few tens of $\mu$eV range has to be used to obtain most of the anti-crossings in $\sigma-$ polarization on the high energy side of the spectra. Depending on relative values of D$_0$ and $E$, splitting of the lines can also be observed at zero magnetic field in some of the dots \cite{Supplemental}.

With a ferromagnetic coupling, at low T the hole-Cr$^+$ is in the ground states with parallel spins configuration and total angular momentum $M_z=\pm 4$. This configuration blocks the optical recombination of the excess 3$d$ electron that would return the Cr atom to its Cr$^{2+}$ ground state. In analogy with dark excitons, this transition is indeed forbidden for parallel hole and 3$d$ electrons spins ({\it i.e} parallel hole and Cr$^+$ spin). The confined hole-Cr$^+$ complex is a particularly favourable system to stabilize the Cr$^+$ as the Coulomb attraction of the hole by the negative charge of the atom enhances their overlap and exchange interaction.

\begin{figure}[hbt]
\centering
\includegraphics[width=1.0\linewidth]{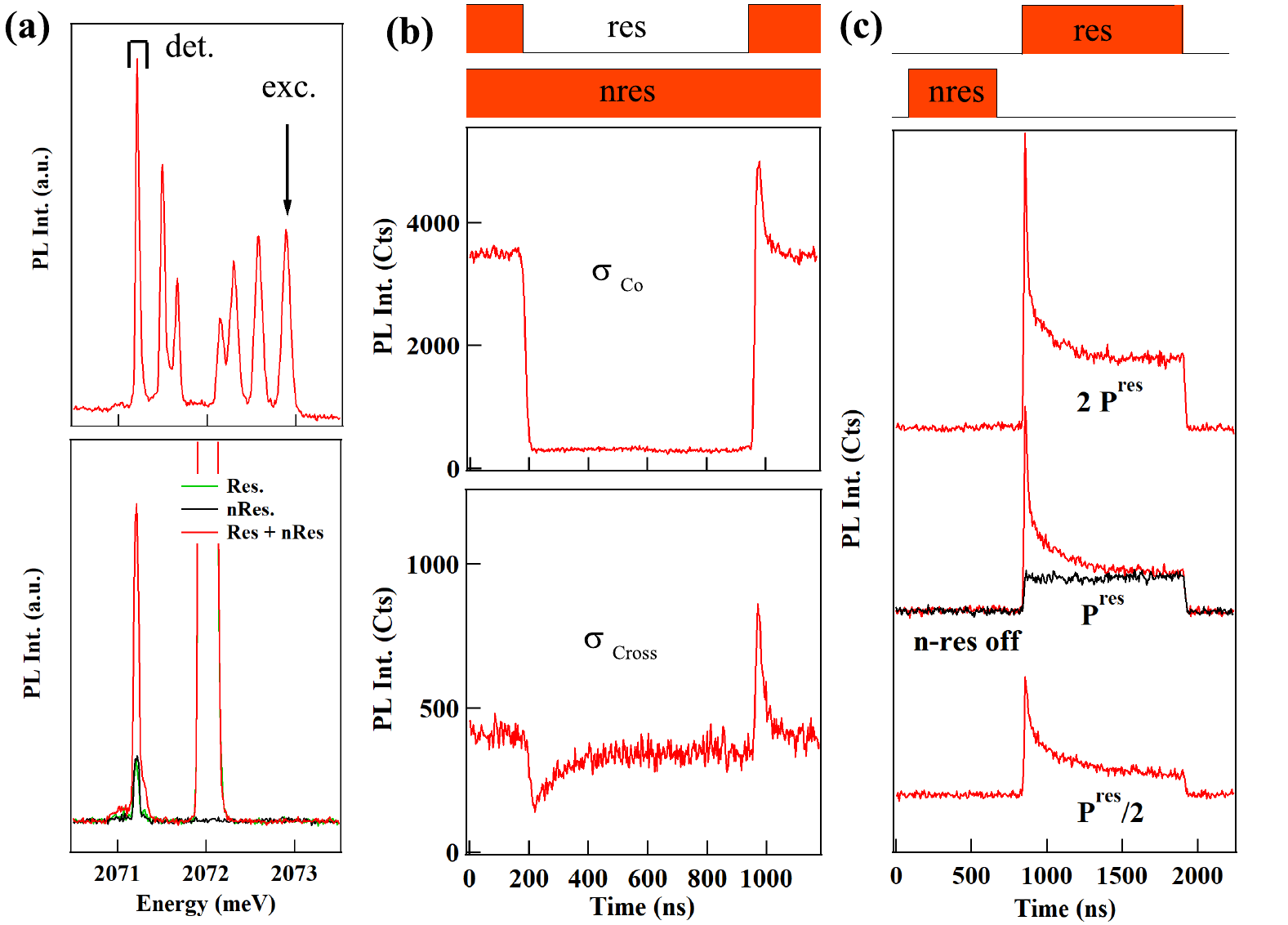}
\caption{(a) Configuration of excitation/detection for pumping experiments (top) and resonant PL observed on the low energy line for an excitation on the hight energy line (bottom). (b) Time resolved resonant PL obtained under continuous non-resonant (568 nm) and pulsed resonant excitations at B=0T. (c) Evolution of the pumping transients as a function of the resonant excitation power in a 2 pulses experiment with co-circular excitation/detection.}
\label{Fig3}
\end{figure}

The two ground hole-Cr$^+$ states $M_z=\pm 4$ are not sensitive to VBM that could induce spin flip-flops as observed in the case of hole-Mn$^{2+}$ \cite{Lafuente2017}. This should enhance the spin memory of the hole-Cr$^+$ nano-magnet. The spin dynamics of hole-Cr$^+$ has been investigated by resonant optical pumping. For a cross-linear excitation and detection, a resonant excitation of the high energy side of X$^+$-Cr$^+$ produces only a weak resonant PL on the low energy line (Fig.~\ref{Fig3}(a)). The resonant PL can be significantly enhanced when an additional non-resonant laser tuned below the ZnTe barrier, on the QD's excited states range, is added. The non-resonant laser produces a weak PL but has a strong influence on the intensity of the resonant fluorescence (Fig.~\ref{Fig3}(a)). 

This behaviour can be explained by the presence of a resonant optical pumping of the hole-Cr$^+$ spin which is partially suppressed by the non-resonant excitation. When the non-resonant excitation is combined with a pulsed resonant excitation on the high energy line, optical pumping transients are observed (Fig.~\ref{Fig3}(b)). The resonant PL is mainly co-circularly polarized with the excitation and a transient is obtained in the emission of the low energy line. In Fig.~\ref{Fig3}(b) the resonant fluorescence obtained for co-circular excitation/detection is compared with the weaker cross-circular signal. In cross-circular configuration, the optical pumping transient is also observed in the resonant fluorescence of the low energy line. In addition, a transient is observed in the non-resonant signal just after the end of the resonant pulse. This is the direct observation in the time domain of the destruction of the pumping by the non-resonant excitation.

Under non-resonant excitation high energy phonons are generated which can contribute to a direct heating of the Cr spin \cite{Tiwari2020Cr}. X$^+$-Cr$^+$ is also formed independently of the hole-Cr$^+$ spin state. Within this complex the Cr$^+$ states are coupled by $E$ (eventually by $I_{eCr^+}$) and spin-flips can occur. Both mechanisms destroy the pumping and the resonant PL is partially restored. The spin dynamics within X$^+$Cr$^+$ is also at the origin of the optical pumping under spin selective excitation. The larger co-polarized resonant PL suggests that spin-flips of the Cr$^+$ are faster than spin-flips of the electron.

\begin{figure}[hbt]
\centering
\includegraphics[width=1.0\linewidth]{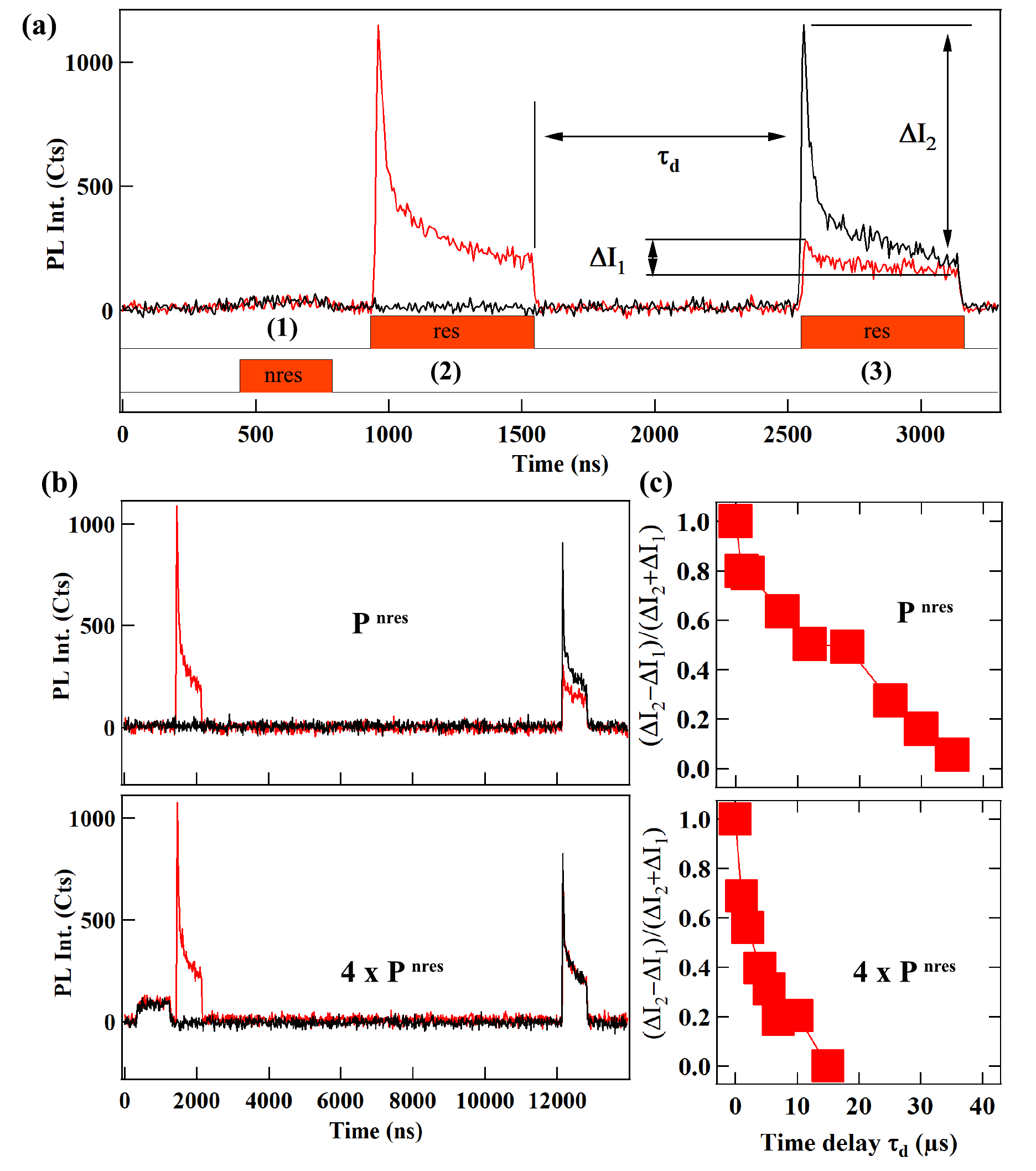}
\caption{(a) Three pulses resonant optical pumping experiment at B=0T in co-circular excitation/detection configuration. (b) Hole-Cr$^+$ spin relaxation observed for a fixed $\tau_d$ and two non-resonant excitation powers P$^{nres}$ and 4$\times$P$^{nres}$. (c) Corresponding measurements of the relaxation time.}
\label{Fig4}
\end{figure}

When the non-resonant and resonant excitations are both modulated and separated in the time domain, an optical pumping is observed in the resonant fluorescence in co-circular excitation/detection configuration (Fig.~\ref{Fig3}(c)). The PL intensity during the resonant pulse reaches a weak intensity plateau after a large transient. The transient is suppressed when the non-resonant excitation is switched off. Its amplitude is much larger than in the presence of the continuous non-resonant excitation showing the increase of the efficiency of the pumping. The dynamics of the optical pumping depends on the power of the resonant laser and can take place in a few tens of ns at high power. It is controlled by the probability of presence of $X^+$ and by the spin dynamics within the X$^+$-Cr$^+$ complex.

To measure the hole-Cr$^+$ spin relaxation we use a two-wavelength time resolved pumping experiment (Fig.~\ref{Fig4}). A non-resonant pulse (pulse 1) is used to initially populates the different hole-Cr$^+$ spin states. A second circularly polarized resonant pulse (pulse 2), tuned to the high-energy line of the QD, is used to perform the optical pumping (i.e., empty the hole-Cr$^+$ spin state under excitation). We use the co-circular excitation / detection configuration where the largest resonant fluorescence signal is obtained. In this configuration, the transient reflects the decrease of the absorption of the QD during the pumping process.

The hole-Cr$^+$ relaxation is probed with a third resonant pulse (pulse 3) with the same energy and polarization as pulse 2 sent after a variable dark time $\tau_d$. The amplitude of the transient during pulse 3 ($\Delta$I$_1$) depends on how the low-energy hole-Cr$^+$ spin state has been populated during $\tau_d$. Alternatively, pulse 2 can be suppressed and the amplitude of the pumping signal obtained during pulse 3 ($\Delta$I$_2$) used as a reference signal (i.e. transient in the absence of initial pumping by pulse 2).

The measured spin relaxation time is presented in Fig.~\ref{Fig4}(b) and (c). The dependence on $\tau_d$ of the normalized amplitude of transients ($\Delta I_2$-$\Delta I_1$)/($\Delta I_2$+$\Delta I_1$) is shown for two values of the intensity of pulse 1. At low non-resonant excitation power a relaxation time around 20 $\mu s$ can be deduced from the time evolution of the pumping transients. The relaxation time decreases with the increase of the power of the non-resonant pulse 1. As recently observed for Cr$^{2+}$, this dependence is likely to be due to the effect of phonons optically generated by the high energy excitation that remain in the sample after the end of the pulse\cite{Tiwari2020Cr}.

The spin relaxation of the hole-Cr$^+$ is more than one order of magnitude longer than what was reported for the hole-Mn$^{2+}$ complex where VBM efficiently couples the spin levels\cite{Lafuente2017}. VBM is still present in the case of hole-Cr$^{+}$ but in the ground states (i.e. M$_z$=$\pm4$ with parallel hole and Cr spins), the ferromagnetic coupling of the hole and Cr$^+$ spin block the hole-Cr$^+$ flip-flops. The hole-Cr$^+$ spin lifetime is then limited by the spin-lattice coupling that cannot be suppressed.

To conclude, we studied a nano-magnet based on a Cr$^+$-doped II-VI semiconductor QD. We demonstrated that the excited state of the Cr, the Cr$^+$ ionized acceptor, can be stable when inserted in a QD and exchanged coupled with a confined heavy-hole spin. The Cr$^+$ spin, not studied until now, is characterised by a ferromagnetic coupling with the hole spin and a vanishingly small exchange interaction with the electron spin. This contrast with Cr$^{2+}$ where an anti-ferromagnetic exchange interaction with the hole spin was observed \cite{Lafuente2018,Besombes2019}. The two hole-Cr$^+$ ground states with angular momentum M$_z=\pm$4 are not sensitive to flip-flops induced by VBM. The measured spin memory at T=4.2 K and zero magnetic field is in the 20 $\mu s$ range. It is limited by the interaction with phonons and could be further improved at sub-Kelvin temperature and weak optical excitation. The presence of Cr$^+$ could be controlled by intentional modulation doping and the hole-Cr$^+$ nano-magnet used as an efficient spin filter in transport devices operating at zero magnetic field and in the 1-2 K range.

\begin{acknowledgements}{}

The work was supported by the French ANR project MechaSpin (ANR-17-CE24-0024). V.T. acknowledges support from EU Marie Curie grant No 754303. BigDFT calculations were done using the French supercomputers GENCI through project 6107. Work in Tsukuba was supported by the Grants-in-Aid for Challenging Exploratory Research (20K21116), Fostering Joint International Research (20KK0113) and JSPS Bilateral Joint Research Project (20219904).

\end{acknowledgements}

\onecolumngrid

\newpage

\begin{center}

{\large{\bf Supplemental Material to \\ "A hole-Cr$^{+}$ nano-magnet in a semiconductor quantum dot"}}

\end{center}

\section{Experimental techniques.}

Individual QDs were studied by optical micro-spectroscopy at liquid helium temperature (T=4.2 K). Piezoelectric actuators and scanners were used to move the sample in front of a high numerical aperture microscope objective (NA=0.85). A magnetic field, up to 9 T along the growth axis of the QDs and 2 T in the plane of the dots, could be applied with a vectorial superconducting coil.

For the magneto-optic measurements, the PL of individual QDs was excited with a continuous wave dye laser tuned to an excited state of the dot, dispersed and filtered by a 1 $m$ double spectrometer before being detected by a Si cooled multichannel charged coupled device (CCD) camera.

For time resolved optical pumping experiments, a dye laser was tuned on resonance with one of the optical transition of X$^+$-Cr$^+$. The laser power was stabilized by an electro-optic variable attenuator. A second diode pumped solid state laser at 568 nm was used for a high energy non-resonant excitation. Both lasers could be modulated by acousto-optic modulators with a time resolution of about 10 ns. A Si avalanche photodiode (APD) with a time resolution of about 350 ps was used in conjunction with a time correlated photon counting system for the time resolved detection.

\section{DFT calculations technique.}

The numerical investigations have been done using the density functional theory within the local density approximation (LDA) exchange and correlation functional. The exact positions of the conduction band and of the top of the valence band is not accurately described by the LDA. This leads to an under-estimation of the band gap of CdTe~\cite{Oba2008}, like for other semi-conductor materials~\cite{Perdew1985}. While this significantly affects any electronic transfer between a defect level and any of the bands (donor or acceptor characters are not quantitative), the respective positions between two defect levels in the band gap are less impacted~\cite{Alkauskas2008}. 

To demonstrate the capacity of a defect to donate an electron to a Cr in substitution, we have studied several common defects in CdTe alloys: anti-sites for both sub-lattices, vacancies on both sub-lattices, [110] split Te-Te interstitial on a Te site and Cd interstitial tetrahedrally coordinated by four Cd atoms. All these structures have been obtained from geometry relaxation runs using the fast inertial relaxation engine (FIRE) in 216-atom super-cells of CdTe in $F\bar{4}3m$ space group. The valence electrons of Cd (12~$e^-$, including semi-cores), Te(6~$e^-$), and Cr(14~$e^-$) were described on a wavelet basis set using BigDFT code\cite{Mohr2015}. The core electrons were frozen in the approximation of pseudo-potentials using the Hartwigsen-Goedecker-Hutter formulation~\cite{Hartwigsen1998}. The calculated value of the CdTe lattice constant for these pseudo-potentials, is 6.42~\AA, which is close to the experimental value of 6.48~\AA.

\begin{figure}[h!]
\centering
\includegraphics[width=0.9\textwidth]{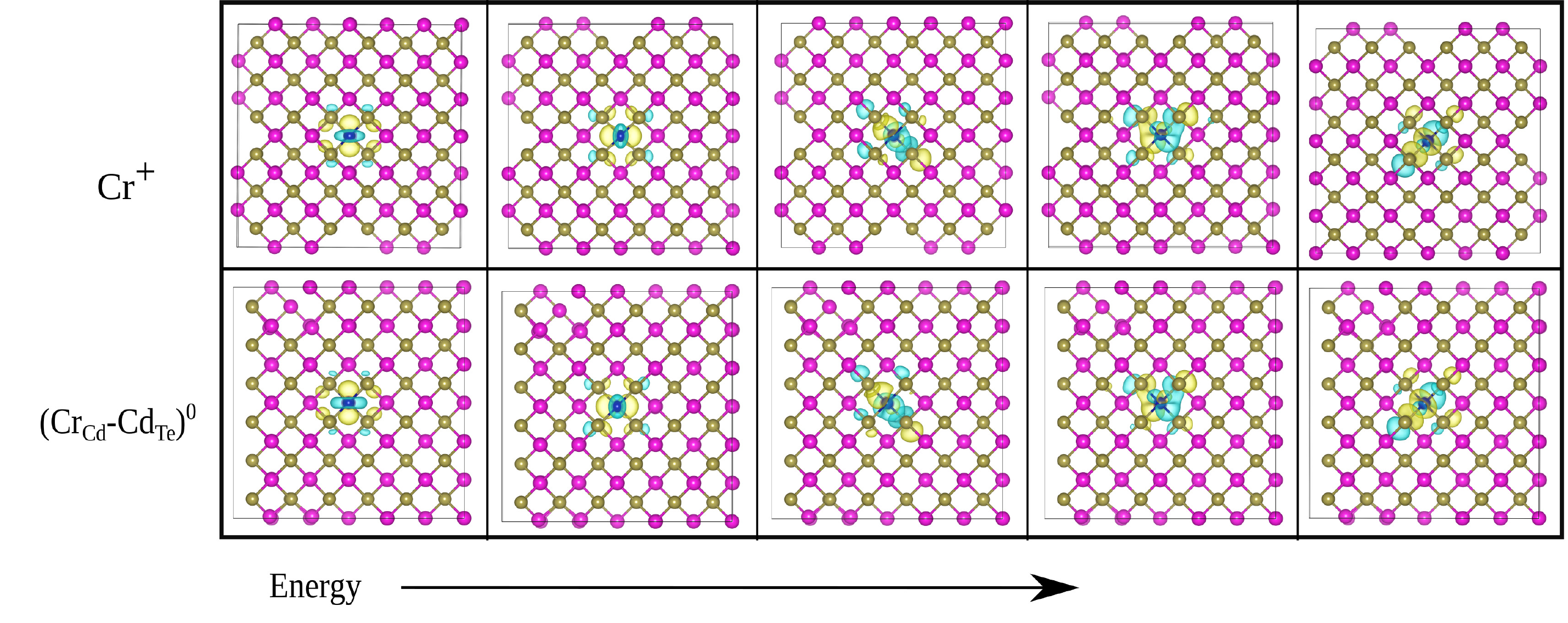}
\caption{Localised states of the neutral system (defect-Cr) in comparison with the same 5 localised states in case of a substitutional Cr with an additional negative charge (Cr$^+$). The presence of the Cd anti-site at 9.5~$\AA$ is providing the electron to the Cr, while not affecting significantly the shape of the KS orbitals. The iso-surface value used for these plots is 0.0025 $e /\AA^3$.}
\label{KS}
\end{figure}

Defects were simulated isolated in the super-cell, but also in the presence of a substitutional Cr as first neighbour. From all tested defects, only the Cd anti-site is presenting a strong donor character. We checked that this donor character is preserved if the Cr is separated from the anti-site defect. In the situation where a Cr is positioned in the same super-cell as the Cd anti-site at a distance of 9.5~$\AA$ (the largest separation in a 216 site super-cell), the electron transfer is visible on the density of state, as shown in Fig.~1 of the letter. In this figure, the position of the Kohn-Sham (KS) eigenvalues in the band gap are depicted with elongated triangles for better readability. While the width of the triangles bears no meaning, the height is representing the degeneracy of the KS states. Tics along the y-axis correspond to a single, double or triple degeneracy. The shape of the KS wave function depicted in Fig.~1 of the letter, representing the p-d hybridization in the case of a Cr$^+$ configuration, is well preserved in the super-cell with the Cd anti-site. As presented in Fig.~\ref{KS}, this is also the case for all the 5 localised levels of the substitutional Cr.

\section{The charge states of Cr in CdTe.}

We summarize here the main experimental results on the possible charge states of Cr in CdTe available in the literature. 

Chromium is incorporated in intrinsic CdTe as the iso-electronic Cr$^{2+}$ impurity but it is also an acceptor. According to references \cite{Godlewski1980} and \cite{Cieplak1975}, the Cr-associated acceptor level (Cr$^{2+}$/Cr$^{+}$) is located within the band-gap and the corresponding optical transition has been identified at approximately E$_{opt}\approx$+1.5 eV above the top of the valence band. The donor level (Cr$^{2+}$/Cr$^{3+}$) is resonant with the valence band states and Cr$^{3+}$ cannot be observed in bulk CdTe. Cr$^{2+}$ is a deep acceptor level and it is not activated at low temperature. However, in the presence of electrical dopant it may captures an electron. Its electronic configuration becomes 3$d^5$ and it acquires a charge -1 with respect to the lattice which characterizes a ionized acceptor.

\begin{figure}[h!]
\centering
\includegraphics[width=0.6\textwidth]{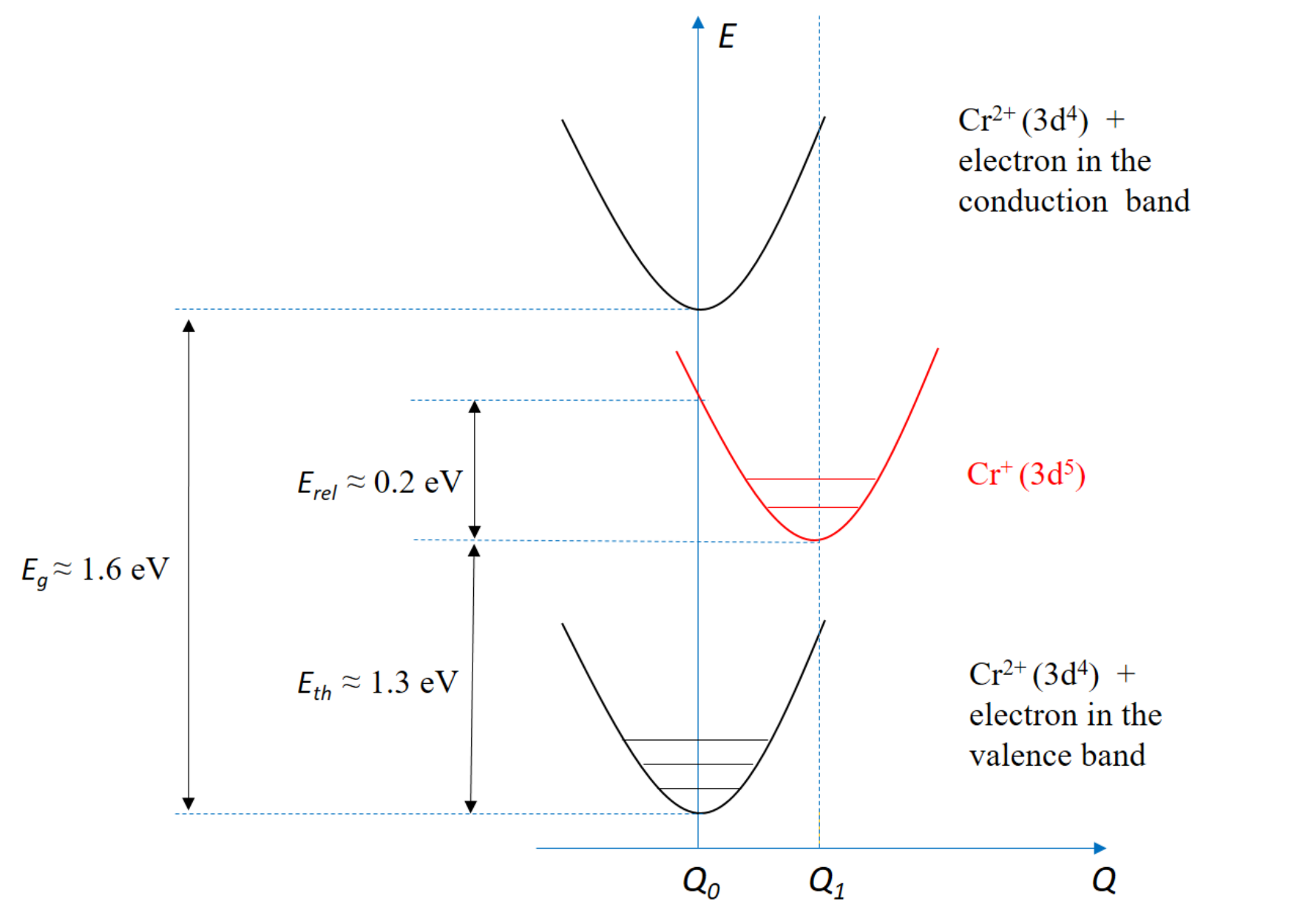}
\caption{Configuration coordinate diagram for Cr$^{2+}$ (3d$^4$) and Cr$^{+}$ (3d$^5$) charge states in CdTe. Q$_0$ and Q$_1$ are the Cr$^{2+}$-Te distance and  Cr$^{+}$-Te distance in the crystal lattice, respectively.}
\label{config}
\end{figure}

As illustrated in Fig.~\ref{config}, the change of the charge state of the ion modifies the impurity-ligand Coulomb interaction and a distortion of the lattice occurs. This distortion stabilizes the Cr$^+$ level. An activation energy around 0.2 eV was deduced from temperature dependent measurements and Cr$^+$ is located at E$_{th}\approx$+1.34 eV above the valence band for a thermal excitation \cite{Cieplak1975}. For a Cr$^+$ in a QD, the distortion of the lattice could be affected by the biaxial strain and this activation energy slightly modified.

The distortion of the lattice reduces the probability of optical recombination of an electron from the 3$d$ orbital with a hole of the valence band and increases the lifetime in the Cr$^+$ excited state. A lifetime in the 1 s range at 100 K has been measured for Cr$^+$ in CdTe in the dark \cite{Godlewski1980} and a longer lifetime could be expected at lower temperature. In a bulk material, this lifetime is significantly shortened under optical excitation where free holes are created \cite{Godlewski1980,Cieplak1975}. 

When located in a QD, the negatively charged Cr$^+$ impurity attracts a free hole. One could expect that the additional electron localized on the 3$d$ shell recombines with the hole. Though inefficient because of the change of local bonds configuration (see Fig.~\ref{config}), this transition would return the Cr atom to its Cr$^{2+}$ ground state. As the electron spin is conserved during the optical transition, this transition is forbidden for parallel hole and 3$d$ electrons spins ({\it i.e} parallel hole and Cr spin). The ferromagnetic exchange interaction with the confined hole prevents the optical recombination of the excess 3$d$ electron and stabilizes the Cr$^+$ ionized acceptor. Under optical excitation, when an electron-hole pair is injected in the dot, the additional hole with anti-parallel spin does not have time to recombine with the excess electron during the lifetime of the positively charged exciton which is around 200 ps.

Let us also note that fluctuations of Cr charge states have been observed for Cr atoms located in the ZnTe barriers in the vicinity of the dots \cite{Besombes2019fluc}. In this case, the exchange interaction with holes in the bulk is weak and parallel spin configuration is not stabilized. The additional 3$d$ electron can recombine with a free hole and fluctuations of the Cr charge state are expected under optical excitation. The situation could be similar for a Cr atom located inside a QD but far from the center of the dot and having a weak exchange interaction with the confined hole. In this case the charge of the Cr atom may fluctuates during the PL acquisition time (typically 1 s) making difficult the detection of the magnetic atom.

\section{Spin properties of Cr$^+$ in CdTe.}

Only the magnetic properties of the 3$d^4$ configuration of the Cr (Cr$^{2+}$) were studied until now in II-VI semiconductors. We present here some elements to understand the coupling of the spin S=5/2 of Cr$^+$ with the bands of the host semiconductor and in particular the sign of the hole-Cr$^+$ exchange interaction. This discussion is mainly based on the theoretical results of reference \cite{Kacman2001}.

\subsection*{Exchange interaction of the electron with Cr$^+$:}

To properly describe the emission of $X^+-Cr^+$ an electron-Cr exchange interaction lower than a few $\mu eV$ has to be used in the model. In diluted magnetic semiconductors, the electron-magnetic atom coupling for carriers at the center of the Brillouin zone mainly arises from the standard short range exchange interaction. It is a ferromagnetic contact interaction which depends on the overlap of the carrier and magnetic atom. The negative localized charge on the Cr$^+$ ion repels the confined electron. This should reduce the overlap between the electron trapped in the dot and the 3$d$ electrons of the Cr$^+$ thus weakening their exchange interaction.

Let us note also that the confinement of the electron reduces the symmetry of its wave function and allows the kinetic $s-d$ exchange to take place. This could give rise to an anti-ferromagnetic contribution to the electron-Cr exchange interaction \cite{Beaulac2010}. This kinetic exchange is however expected to be very weak and it is most likely the reduction of the overlap which explains the weak electron-Cr$^+$ exchange interaction. A similar behaviour is observed in III-V QDs where a negligible electron-(Mn$^{2+}$+h) exchange interaction was needed to explain the spectra a InAs/GaAs QDs doped with a single Mn atom \cite{Krebs2009}.

\subsection*{Exchange interaction of the hole with Cr$^+$:}

An anti-ferromagnetic coupling with a confined hole was observed for Cr$^{2+}$ in CdTe/ZnTe QDs \cite{Lafuente2018}. The most studied 3$d^5$ element, Mn$^{2+}$, also presents an anti-ferromagnetic coupling with holes in CdTe. The exchange coupling of the spin 5/2 of a Cr$^+$ with carriers' spins in II-VI compounds have never been studied until now. We present in this work the first measurement of the hole-Cr$^+$ exchange interaction in CdTe and give here some elements to understand its ferromagnetic sign. 

In diluted magnetic semiconductors, the exchange interaction of the localized spins of magnetic atoms with the hole spin is usually larger than the interaction with the electron spin and arises from two mechanisms: (i) a ferromagnetic coupling resulting from the short range exchange interaction and (ii) a spin dependent hybridization of the 3$d$ orbital of the magnetic atom and the $p$ orbital of the host semiconductor, the so-called kinetic exchange.

The $p-d$ hybridization is strongly sensitive to the energy splitting between the 3$d$ levels of the atom and the top of the valence band. This hybridization of course significantly depends on the considered transition metal element (i.e. filling of the $3d$ orbital) and, for a given magnetic element, on the semiconductor host \cite{Kossut}. The resulting kinetic exchange can be either ferromagnetic or anti-ferromagnetic depending on the relative position of the $d$ levels and the top of the valence.

In the case of Mn in II-VI compounds, the Mn$^{2+/3+}$ donor level is located far within the valence band and the kinetic exchange results in an anti-ferromagnetic hole-Mn coupling. The p-d hybridization has the main contribution to the hole-Mn exchange and the overall interaction is anti-ferromagnetic \cite{Furdyna1988}.

For a Cr$^+$ ion in the 3$d^5$ configuration, the exchange interaction with a heavy-hole with a $z$ component of its total angular momentum J$_z$=$\pm$3/2 can be expressed in the form \cite{Kacman2001}:

\begin{eqnarray}
H_{ex}=-\frac{1}{3}S_zJ_zB_5
\label{exchange}
\end{eqnarray}

\noindent where $B_N$ (with N=5 for Cr$^+$) is given by

\begin{eqnarray}
B_N=-\frac{V_{pd}^2}{S}[\frac{1}{\varepsilon_p+E_{N-1}^{S-1/2}-E_{N}^{S}}+\frac{1}{E_{N+1}^{S-1/2}-E_{N}^{S}-\varepsilon_p}]
\label{BN}
\end{eqnarray}

\noindent with $E_N^S$ the unperturbed energy of the $d$ shell with N electrons and total spin S, V$_{pd}$ the hybridization constant between the $d$ orbital of the impurity and the $p$ orbital of the semiconductor host and $\varepsilon_p$ the energy of the top of the valence band \cite{Kacman2001}. 

In the expression (\ref{BN}) controlling the amplitude of the hole-Cr exchange, the first denominator corresponds to the energy $e_1$ required to transfer an electron from the $d$ shell of the Cr$^+$ ion to the valence band reducing the total spin from S to S-1/2 (and reducing the number of electrons in the $d$ shell from N to N-1). The denominator of the second term is the energy $e_2$ required to transfer an electron from the top of
the valence band to the $d$ shell also with a reduction by 1/2 of the total spin (and an increase of the number of electrons in the $d$ shell from N to N+1). This last energy $e_2$ includes the electron-electron exchange interaction in the $d$ shell (at the origin of Hund rule) and is always positive and very large (a few eV).

The sign of $B_5$ controls the sign of the hole-Cr kinetic exchange interaction: a negative $B_5$ corresponds to an anti-ferromagnetic interaction whereas a positive $B_5$ will give rise to a ferromagnetic interaction. In the case of a Cr$^+$ ion, the donor transition Cr$^{+}$ to Cr$^{2+}$ is within the band gap of CdTe, e$_1$ is negative ($E_{N}^{S}-E_{N-1}^{S-1/2}>\varepsilon_p$), $1/e_1<-1/e_2$ and $B_5$ is positive. The hole-Cr exchange interaction is ferromagnetic as observed in our experiments. This is opposite to the situation of the commonly used diluted magnetic semiconductors based on a doping with the 3$d^5$ Mn$^{2+}$.

Expression \ref{BN} results from a perturbation approach and should be used with care. However, it should be valid for a hole of the top of the valence band ({\it i.e.} not strongly localized on a ionized impurity) and for a large value of $e_1$. These conditions are apparently fulfilled for Cr$^+$ in CdTe.

\section{Valence band mixing in Cr$^+$-doped QDs.}

A description of the optical properties of magnetic QDs usually requires to go beyond the heavy-hole approximation and take into account some effect of valence band mixing (VBM). We present here the details of the pseudo-spin model used in the manuscript to describe VBM.

As discussed in the manuscript, the spin structure of the hole-Cr$^+$ nano-magnet is described by the effective spin Hamiltonian 

\begin{eqnarray}
{\cal H}_{h-Cr^+}=I_{hCr^+}\vec{S}\cdot\vec{J}+g_{h}\mu_B\vec{J}\cdot\vec{B}+{\cal H}_{Cr^+}
\end{eqnarray}

\noindent and VBM will mainly affect the hole-Cr$^+$ energy levels through the exchange coupling term $I_{hCr^+}\vec{S}\cdot\vec{J}$. 

In self-assembled QDs, the VBM mainly arises from strain anisotropy. We can write the Bir-Pikus Hamiltonian describing the influence of strain on the valence band structure in the basis $\lbrace+3/2,+1/2,-1/2,-3/2\rbrace$ as:

\begin{eqnarray}
H_{BP}= 
\begin{pmatrix}
p+q & s   & r    & 0  \\ 
s^* & p-q & 0    & r  \\ 
r^* & 0   & p-q  & -s\\ 
0   & r^* & -s^* & p+q
\end{pmatrix} 
\end{eqnarray}

\noindent with 

\begin{eqnarray}
p=a_v(\epsilon_{xx}+\epsilon_{yy}+\epsilon_{zz});q=b(\epsilon_{zz}-(\epsilon_{xx}+\epsilon_{yy})/2)\\ \nonumber
r=b\frac{\sqrt{3}}{2}(\epsilon_{xx}-\epsilon_{yy})-id\epsilon_{xy};s=d(\epsilon_{xz}-i\epsilon_{yz})
\end{eqnarray}

\noindent where $a_v$, $b$ and $d$ are deformation potential of the valence band and $\epsilon_{ij}$ the strain tensor components. The non-diagonal terms $s$ and $r$ couple the ground heavy-holes states with light-holes. 

For a weak VBM, a pseudo-spin description of the perturbed heavy-holes ground states is usually enough to understand the main effect of the strain induced coupling on the optical properties of QDs.

\begin{figure}[h!]
\centering
\includegraphics[width=0.8\textwidth]{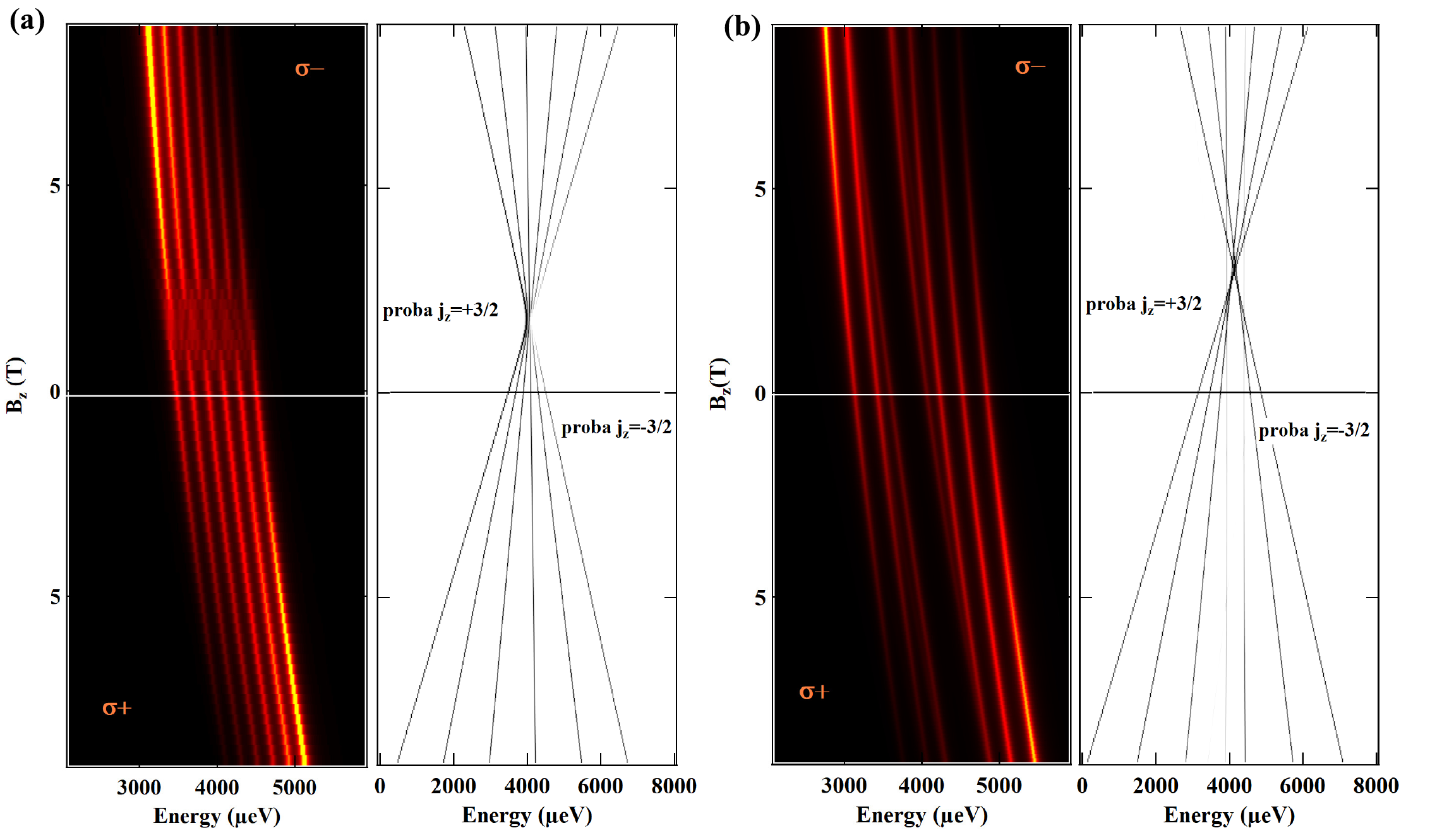}
\caption{(a) Influence of shear strain on the magneto-optic properties of Cr$^+$ doped QDs. Left: Calculated PL spectra. Right: Calculated hole-Cr$^+$ energy levels with $I_{eCr^+}=0\mu eV$, $I_{hCr^+}=-135\mu eV$ (exchange parameters of QD1 of the manuscript), $T_{eff}=20K$, $\rho_s/\Delta_{lh}=0.0$, $\theta_s=-\pi/4$, $\gamma$=1.5 $\mu eV T^{-2}$, g$_{Cr^+}$=2, g$_{h}$=0.5, g$_{e}$=-0.4, E=0 $\mu eV$, D$_0$=0 $\mu eV$ and $\xi=0.15$. (b) Influence of in-plane strain anisotropy on the magneto-optic properties of Cr$^+$ doped QDs. Left: Calculated PL spectra. Right: Calculated hole-Cr$^+$ energy levels with with $I_{eCr^+}=0\mu eV$, $I_{hCr^+}=-225\mu eV$ (exchange parameters of QD2 of the manuscript), $T_{eff}=20K$, $\rho_s/\Delta_{lh}=0.2$, $\theta_s=-\pi/4$, $\gamma$=1.5 $\mu eV T^{-2}$, g$_{Cr^+}$=2, g$_{h}$=0.5, g$_{e}$=-0.4, E=0 $\mu eV$, D$_0$=0 $\mu eV$ and $\xi=0$. }
\label{Fig2S}
\end{figure}

\subsection*{Influence of in-plane strain anisotropy}

In the presence of strain anisotropy in the QD plane ($r \neq 0$), the two hole ground states become:

\begin{eqnarray}
\vert\Phi^+_h\rangle=\vert+3/2\rangle-\varphi\vert-1/2\rangle\nonumber\\
\vert\Phi^-_h\rangle=\vert-3/2\rangle-\varphi^*\vert+1/2\rangle
\end{eqnarray}

\noindent with $\varphi=\rho_s/\Delta_{lh}e^{2i\theta_s}$ describing the amplitude of the mixing and $\Delta_{lh}$ the splitting of light-holes and heavy-holes induced both by strain and confinement. $\theta_s$ is an angle relative to the [100] axis describing the direction of the in-plane strain anisotropy.

A first order development of the angular momentum operator $J$ on the subspace of the perturbed holes $\vert\Phi^{\pm}_h\rangle$ leads to

\begin{eqnarray}
\tilde{j}_+= 
\frac{\rho_s}{\Delta_{lh}}\begin{pmatrix}
0 & -2\sqrt{3}e^{-2i\theta_s} \\ 
0 & 0 
\end{pmatrix}; 
\tilde{j}_-=
\frac{\rho_s}{\Delta_{lh}}\begin{pmatrix}
0 & 0 \\ 
-2\sqrt{3}e^{2i\theta_s} & 0 
\end{pmatrix};
\tilde{j}_z=
\begin{pmatrix}
3/2 & 0 \\ 
0 & -3/2 
\end{pmatrix}
\end{eqnarray}

\noindent $\tilde{j}_+$ and $\tilde{j}_-$ flip the hole spin whereas a measurement of the spin projection along $z$ confirms that they are mainly heavy holes. This type of VBM unlock the spinflips between the hole and its surrounding medium. It allows hole-Cr$^+$ flip-flops and couples the different hole-Cr$^+$ spin states. This is in particular the case for the states $\vert S_z=-1/2,\Uparrow_h\rangle$ and $\vert S_z=+1/2,\Downarrow_h\rangle$ which are mixed and split (Fig.~\ref{Fig2S}(b)). This mixing term is responsible for the opening of the gap in the center of the X$^+$-Cr$^+$ spectra.

\subsection*{Influence of shear strain}

In the presence of shear strain ($s\neq 0$), the two hole ground states become:

\begin{eqnarray}
\vert\Xi^+_h\rangle=\vert+3/2\rangle+\xi\vert+1/2\rangle\nonumber\\
\vert\Xi^-_h\rangle=\vert-3/2\rangle-\xi^*\vert-1/2\rangle
\end{eqnarray}

A first order development of the angular momentum operator $J$ on the subspace of the perturbed holes $\vert\Xi^{\pm}_h\rangle$ leads to

\begin{eqnarray}
\tilde{j}_+= 
\xi\begin{pmatrix}
\sqrt{3} & 0 \\ 
0 & -\sqrt{3} 
\end{pmatrix}; 
\tilde{j}_-=
\xi^*\begin{pmatrix}
\sqrt{3} & 0 \\ 
0 & -\sqrt{3} 
\end{pmatrix};
\tilde{j}_z=
\begin{pmatrix}
3/2 & 0 \\ 
0 & -3/2 
\end{pmatrix}
\end{eqnarray}

Because of this valence band mixing the hole-Cr$^+$ exchange interaction couples the states $\vert\Xi^{\pm}_h,S_z\rangle$ with the states $\vert\Xi^{\pm}_h,S_z+1\rangle$ and $\vert\Xi^{\pm}_h,S_z-1\rangle$. This coupling as an influence under a positive longitudinal magnetic field as the states $\vert\Xi^{+}_h,S_z\rangle$ are crossings. This appends for instance around B$_z$=1T with the exchange parameters of QD1. At this crossing point all the states are mixed $\xi$ and the emission spectra are perturbed. It is difficult to precisely deduce this parameter from the optical spectra. In Fig.~\ref{Fig2S}(a) spectra of QD1 are however quite well described with $\xi\approx0.15$ and $E=0\mu eV$ (we also use $D_0=0\mu eV$ as its value cannot be estimated from the spectra when $E=0\mu eV$).

\section{Additional examples of Cr$^+$-doped QDs.}

The emission spectra of a magnetic QD depends strongly on the position of the magnetic atom inside the dot which controls the exchange interaction with confined carriers and on the strain distribution which controls the VBM. The local strain at the magnetic atom location controls the fine structure of the atom and can also have an effect on the emission spectra.

\begin{figure}[hbt]
\centering
\includegraphics[width=1.0\textwidth]{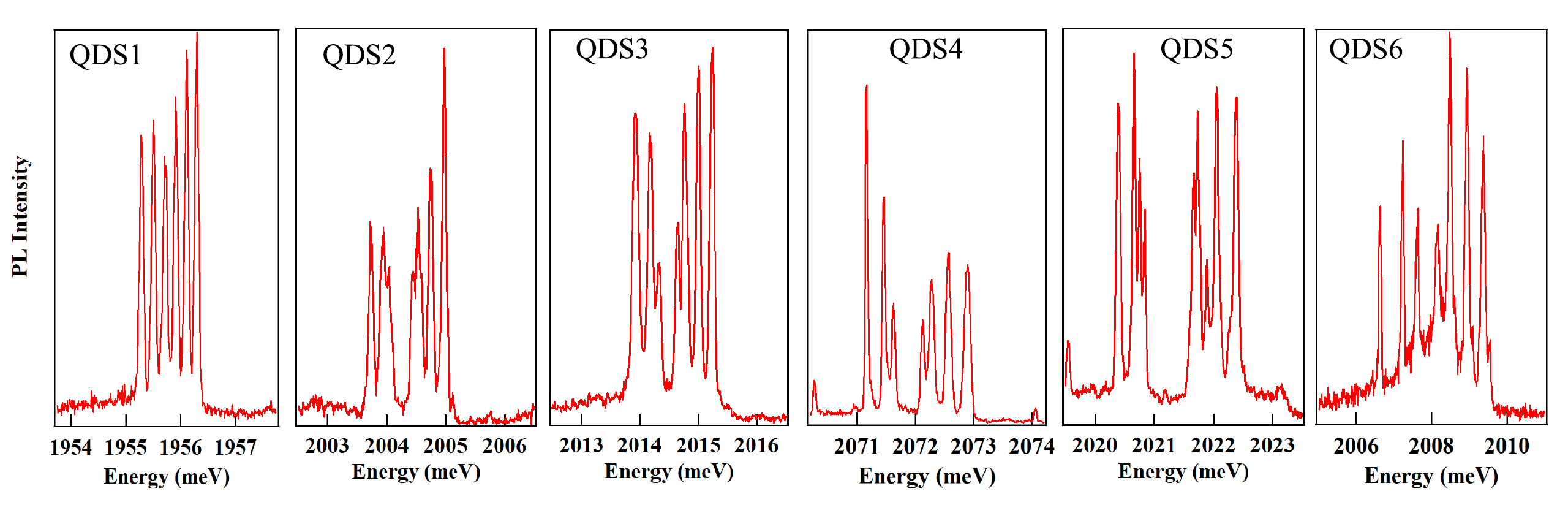}
\caption{Photoluminescence spectra at T=4.2K and at zero magnetic field of six positively charged QDs containing a Cr$^+$ ion. The spectra are ordered with increasing energy splitting (note the difference in the energy scale of QDS6). QDS1 and QDS4 correspond to the original QDs used in the manuscript.}
\label{Fig3S}
\end{figure}

\begin{figure}[h!]
\centering
\includegraphics[width=0.6\textwidth]{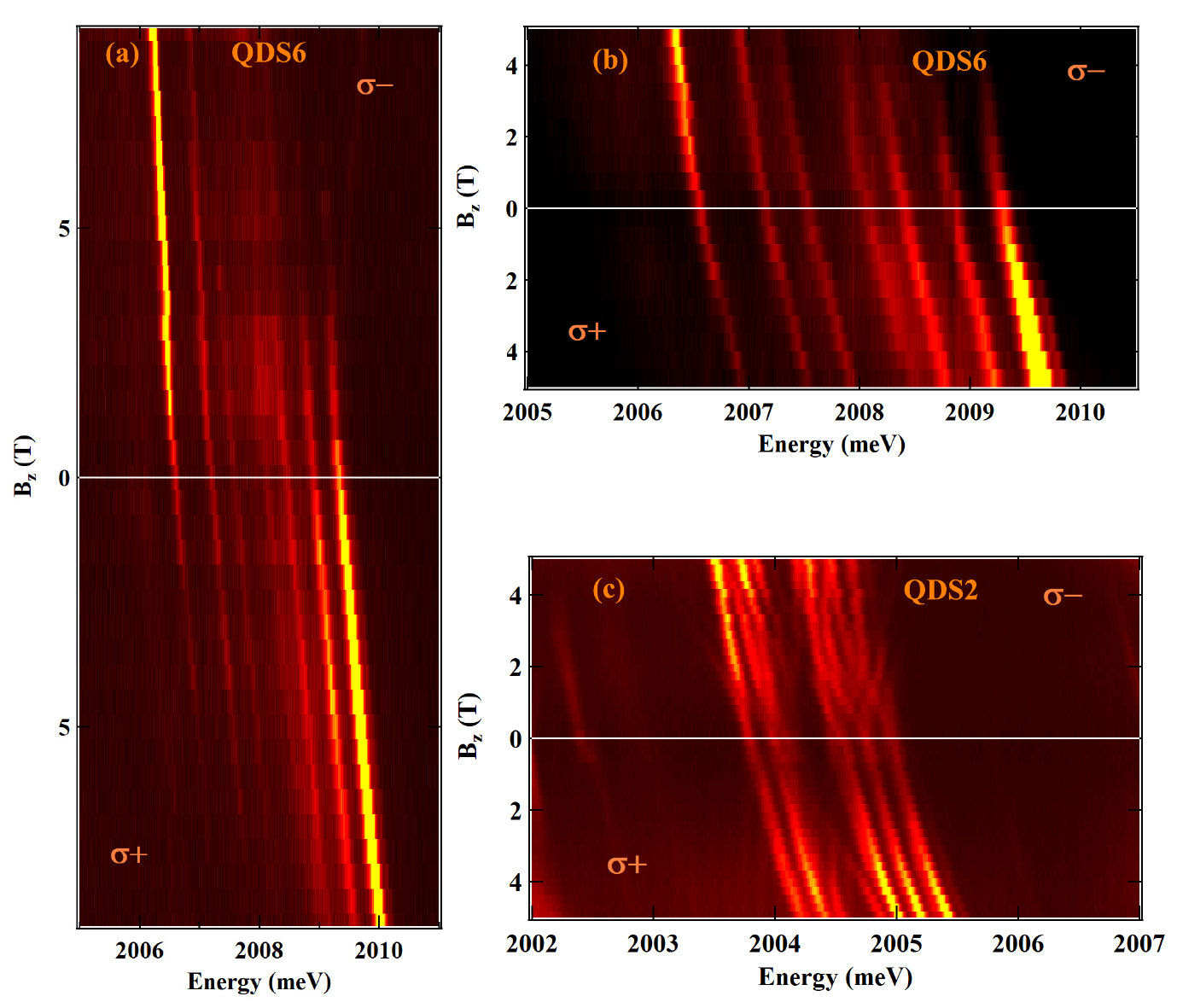}
\caption{Intensity map of the magnetic field dependence of the PL of the dots presenting a large (QDS6 in (a) and the corresponding low magnetic field detail in (b)) and a small (QDS2 in (c)) zero field overall splitting.}
\label{Fig4S}
\end{figure}

\begin{figure}[h!]
\centering
\includegraphics[width=0.8\textwidth]{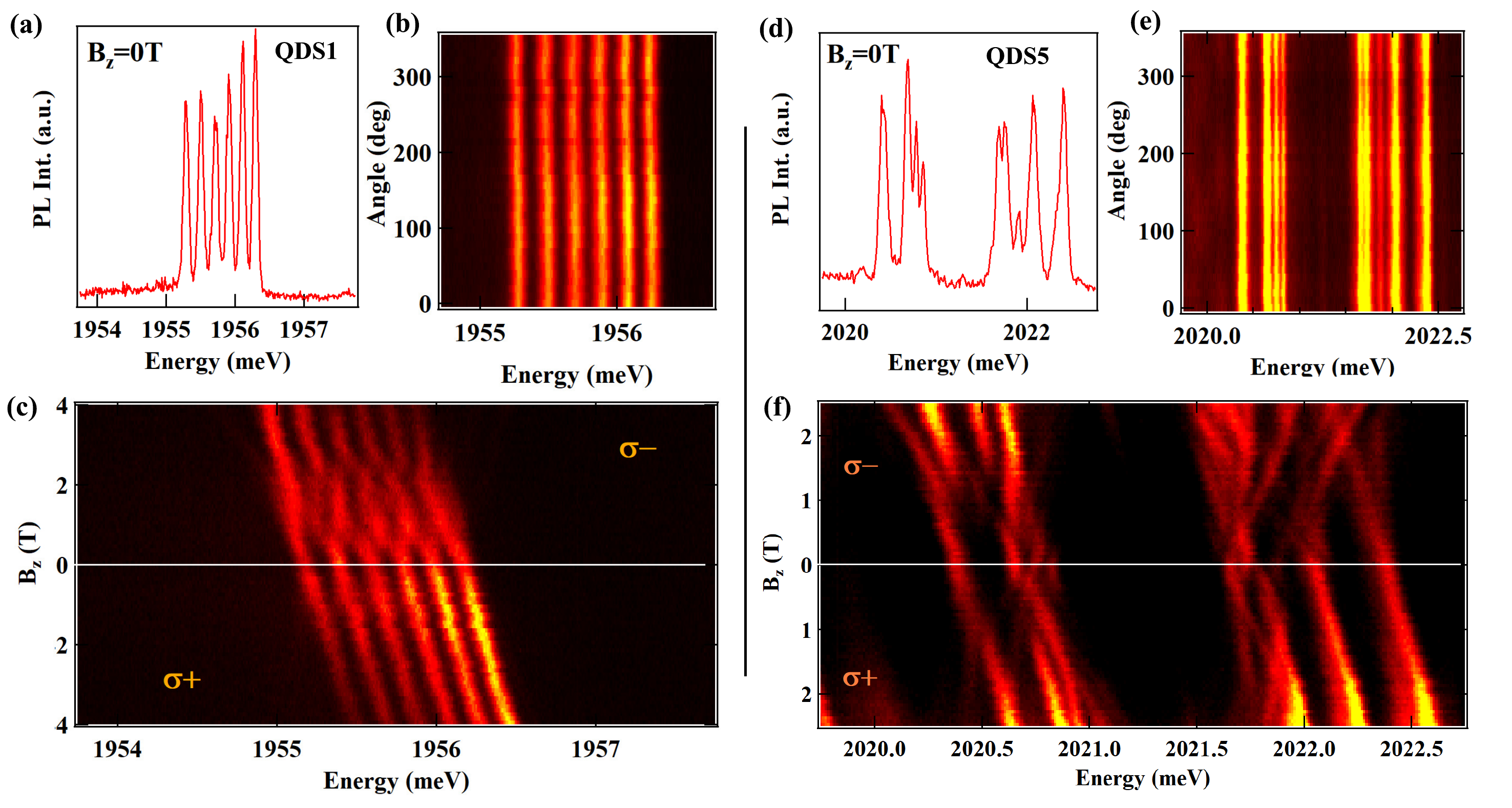}
\caption{Left: Detail at low magnetic field of the PL of a dot with no in-plane strain anisotropy and 6 lines at zero field (QDS1): (a) PL spectra at zero field. (b) Linear polarization PL intensity map at zero magnetic field. (c) Magnetic field dependence of the circularly polarized PL intensity map.
Right: Detail at low magnetic field of the PL of dot with a large in-plane strain anisotropy and more than 7 lines at zero field (QDS5). (d) PL spectra at zero field. (e) Linear polarization PL intensity map at zero magnetic field. (f) Magnetic field dependence of the circularly polarized PL intensity map.}
\label{Fig5S}
\end{figure}

To illustrate this in the case of Cr$^+$ and show the diversity of the observed spectra, we present in figure \ref{Fig3S} the PL at T=4.2K and B=0 T of six QDs containing a hole-Cr$^+$ complex. The overall splitting of the lines as well as the width of the central gap changes from dot to dot. In QDS5 which presents the largest central gap ({\it i.e.} large strain induced valence band mixing) some of the lines are split and more than 7 lines are observed at zero magnetic field. In QDS1 (QD1 of the manuscript) the intensity of the third line is slightly weaker than the others showing that it starts to be split. This dot presents almost no valence band mixing.

All the dots present the same behaviour under a large longitudinal magnetic field with a thermalization on the high energy line in $\sigma+$ polarization and on the low energy line in $\sigma-$ polarization (see figure \ref{Fig4S}). This results from the ferromagnetic hole-Cr$^+$ exchange interaction.

\subsection*{Details of anti-crossings at low magnetic field}

\begin{figure}[h!]
\centering
\includegraphics[width=0.6\textwidth]{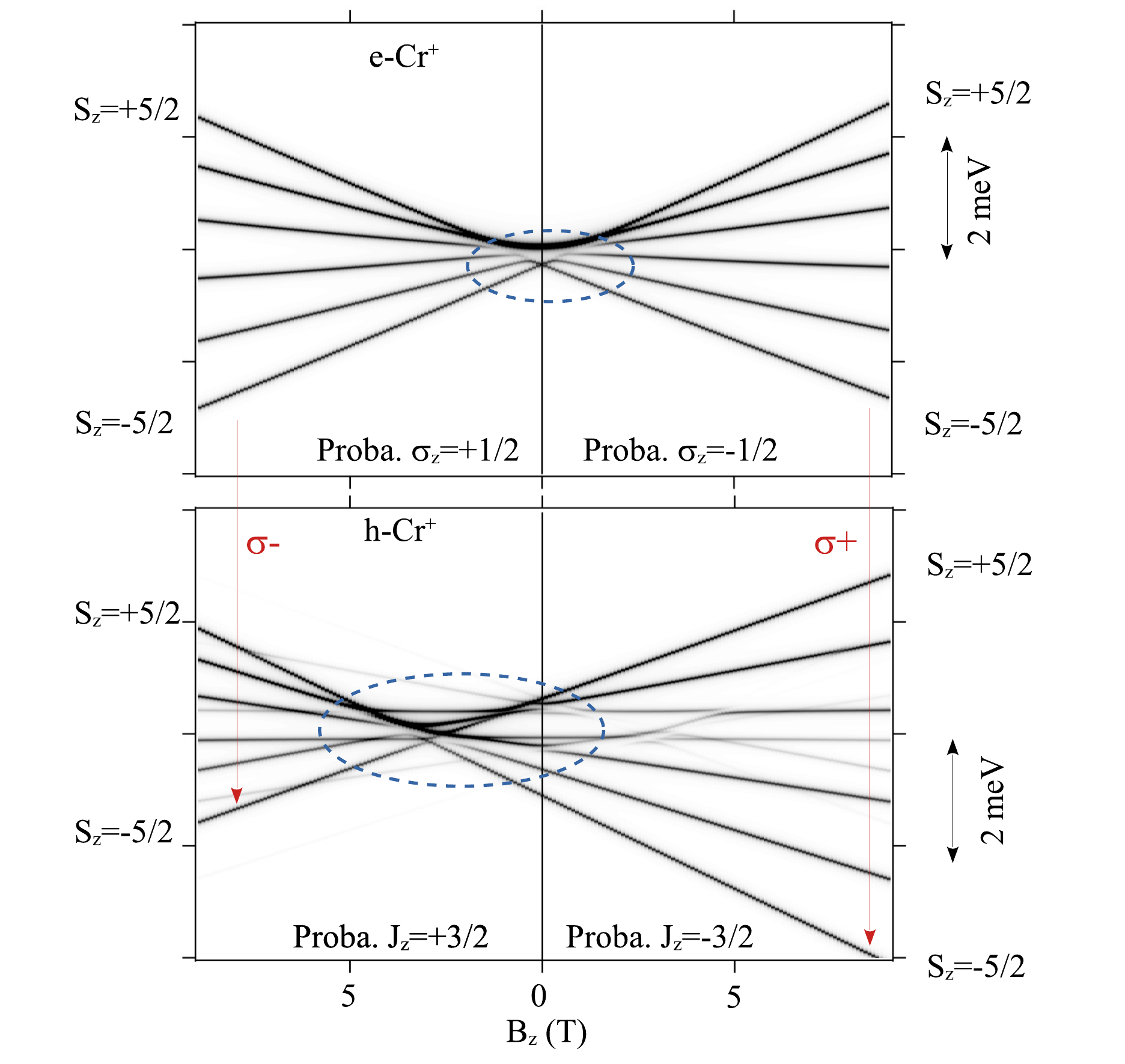}
\caption{Calculated energy levels of electron-Cr$^+$ (i.e. X$^+$-Cr$^+$) and hole-Cr$^+$ as a function of magnetic field. Energy parameters are those of Fig. 3 of the Letter but no thermalisation is used. The most intense transitions in $\sigma+$ and $\sigma-$ polarizations are indicated. Blue dashed lines point out the magnetic field range where the influence of the $E$ term is the most significant.}
\label{fanS}
\end{figure}

\begin{figure}[h!]
\centering
\includegraphics[width=0.8\textwidth]{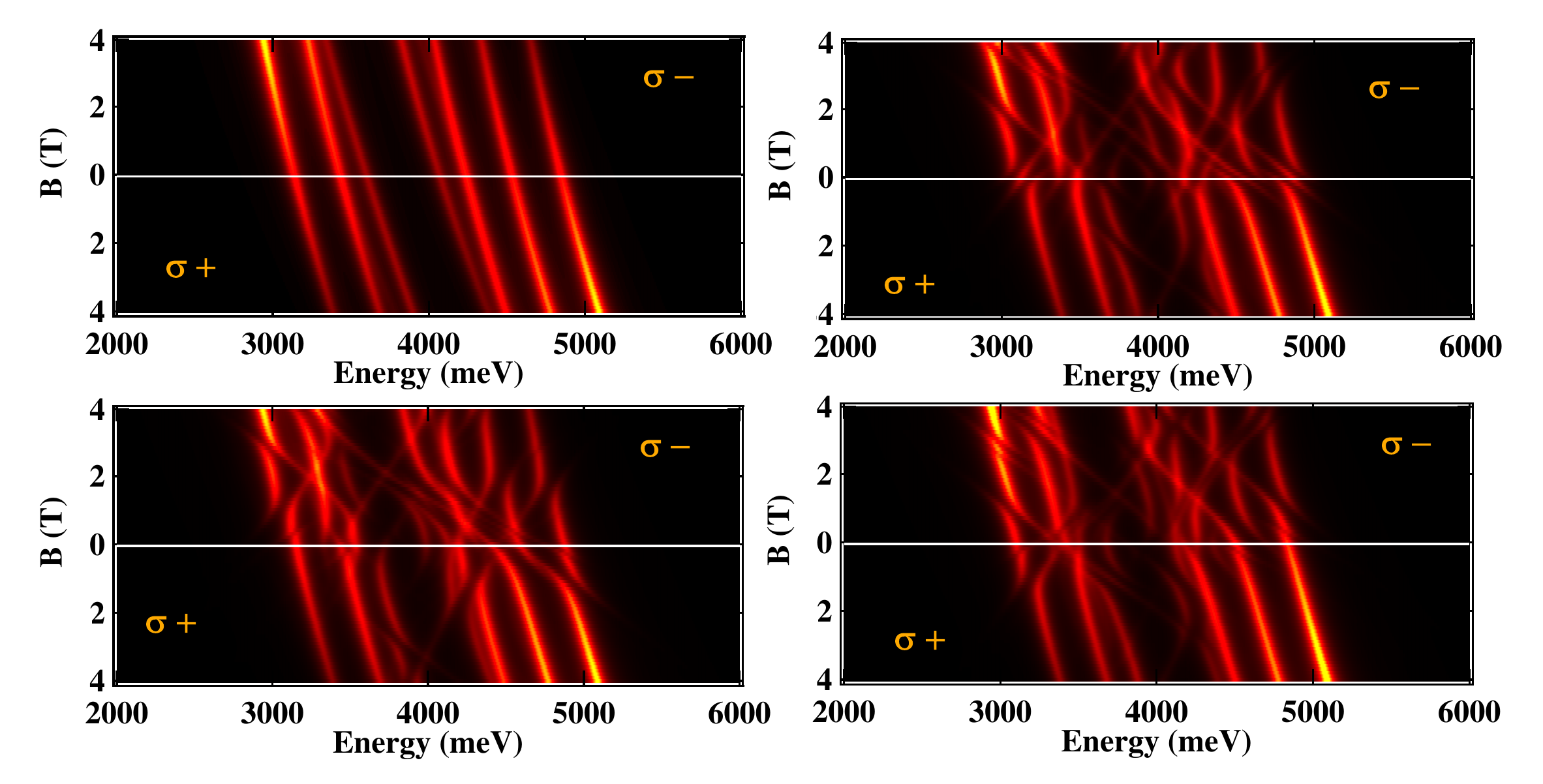}
\caption{Calculated longitudinal magnetic field dependence of X$^+$-Cr$^+$ for $I_{eCr^+}=0\mu eV$, $I_{hCr^+}=-225\mu eV$, $T_{eff}=20K$, $\rho_s/\Delta_{lh}=0.2$, $\theta_s=-\pi/4$, $\gamma$=1.5 $\mu eV T^{-2}$, g$_{Cr^+}$=2, g$_{h}$=0.5, g$_{e}$=-0.4, $\xi\approx0.0$ (parameters of QD2 of the manuscript) and different values of the Cr$^+$ fine structure terms: (a) E=0 $\mu eV$ and D$_0$=0 $\mu eV$, (b) E=20 $\mu eV$ and D$_0$=0 $\mu eV$ (c) E=20$\mu eV$ and D$_0$=+40$\mu eV$ (d) E=20 $\mu eV$ and D$_0$=-40 $\mu eV$.}
\label{Fig6S}
\end{figure}

The observed behaviour at low field can however differ from dot to dot. Most of the dots present a central gap induced by in-plane strain anisotropy. In addition, a complex anti-crossing structure is often observed in the low field region.

Fig.~\ref{fanS} presents the magnetic field dependence of the electron-Cr$^+$ ({\it i.e.} X$^+$-Cr$^+$) and hole-Cr$^+$ energy levels calculated with the parameters of Fig.3 of the Letter. At low magnetic field, in addition to the gap in the hole-Cr$^+$ levels induced by the VBM (see Fig.3 of the Supplemental Materials) electron-Cr$^+$ levels in the excited state and hole-Cr$^+$ levels in the ground state are mixed by the Cr$^+$ fine structure term $E$. The effect is significant around 0 T on the electron-Cr$^+$ levels and in the range 0T-4T for the hole-Cr$^+$ levels in the final state of a $\sigma-$ recombination. This gives rise to the complex spectra presented in Fig. 3 of the Letter where anti-crossings appear mainly in the range 0T-4T in $\sigma-$ polarization.

The width an the position of the anti-crossings depends on the values of the Cr$^+$ fine structure terms $E$ and $D_0$. These terms are controlled by local strain at the Cr atom position. The amplitude of the anti-crossing is governed by $E$ whereas their position under magnetic field are controlled by $D_0$. The influence of both parameters on the low magnetic field dependence of X$^+$-Cr$^+$ are presented in Fig.~\ref{Fig6S} where calculation are performed with the parameters of QD2. Some particular values of these parameters can lead to the observation of line splittings at zero field. This is observed for instance on the low magnetic field PL dependence of QDS5 presented in figure \ref{Fig5S}. In the opposite, QDS6 with the largest exchange induced splitting, does not present any significant strain induced mixing of the Cr$^+$ states.

\subsection*{Transverse magnetic field dependence}

\begin{figure}[h!]
\centering
\includegraphics[width=0.7\textwidth]{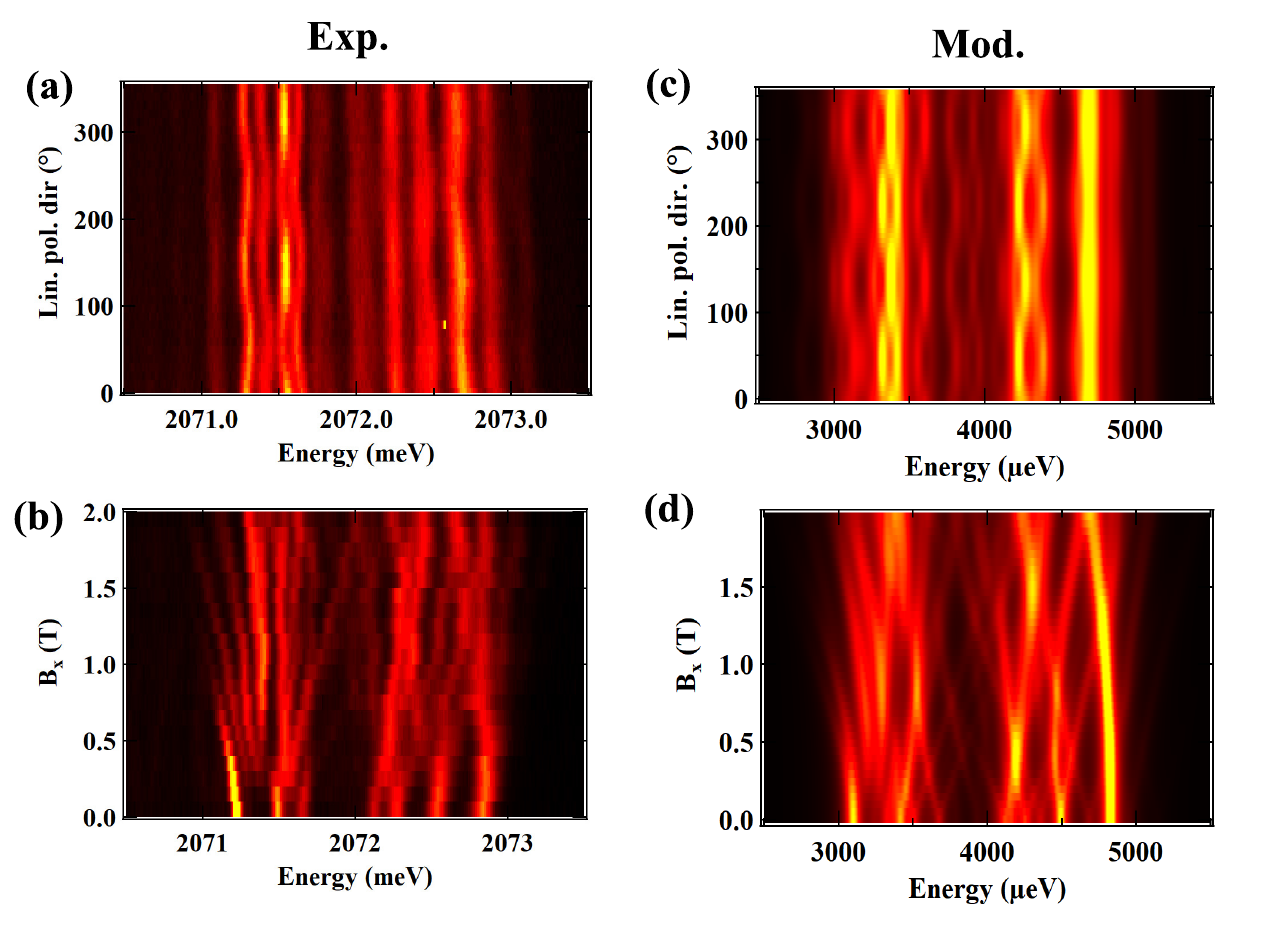}
\caption{Transverse magnetic field dependence (b) and linear polarization under a transverse magnetic field B$_x$=2T (a) of QDS4 (QD2 of the manuscript). (c) and (d): Model of the transverse magnetic field dependence obtained with the parameters used in figure 3 of the manuscript. The transverse magnetic field is along the [100] axis.}
\label{Fig7S}
\end{figure}

Under a weak transverse magnetic field (see figure \ref{Fig7S} in the case of QDS4) a complex PL behaviour is also observed. It results from a progressive alignment of the Cr$^+$ spin along the transverse field when the dot is occupied by a charged exciton ({\it i.e.} electron-Cr$^+$ levels). In the final hole-Cr$^+$ ground state state and in the moderate transverse field range analysed here, the Cr$^+$ spin remains quantized along the QD growth axis as the Zeeman energy of the Cr$^+$ is weaker than I$_{hCr^+}$. With a change of quantization axis during the optical recombination, many transitions becomes progressively allowed as the transverse field is increasing.

As observed for charged excitons in non-magnetic QDs, in a transverse magnetic field the emission becomes linearly polarized. The transverse magnetic field dependence and the linear polarization observed under transverse magnetic field can also be well reproduced by the spin effective model presented in the manuscript confirming the validity of the proposed description of observed spectra.

\section{PL intensity distribution of X$^+$-Cr$^+$.}

\begin{figure}[h!]
\centering
\includegraphics[width=0.5\textwidth]{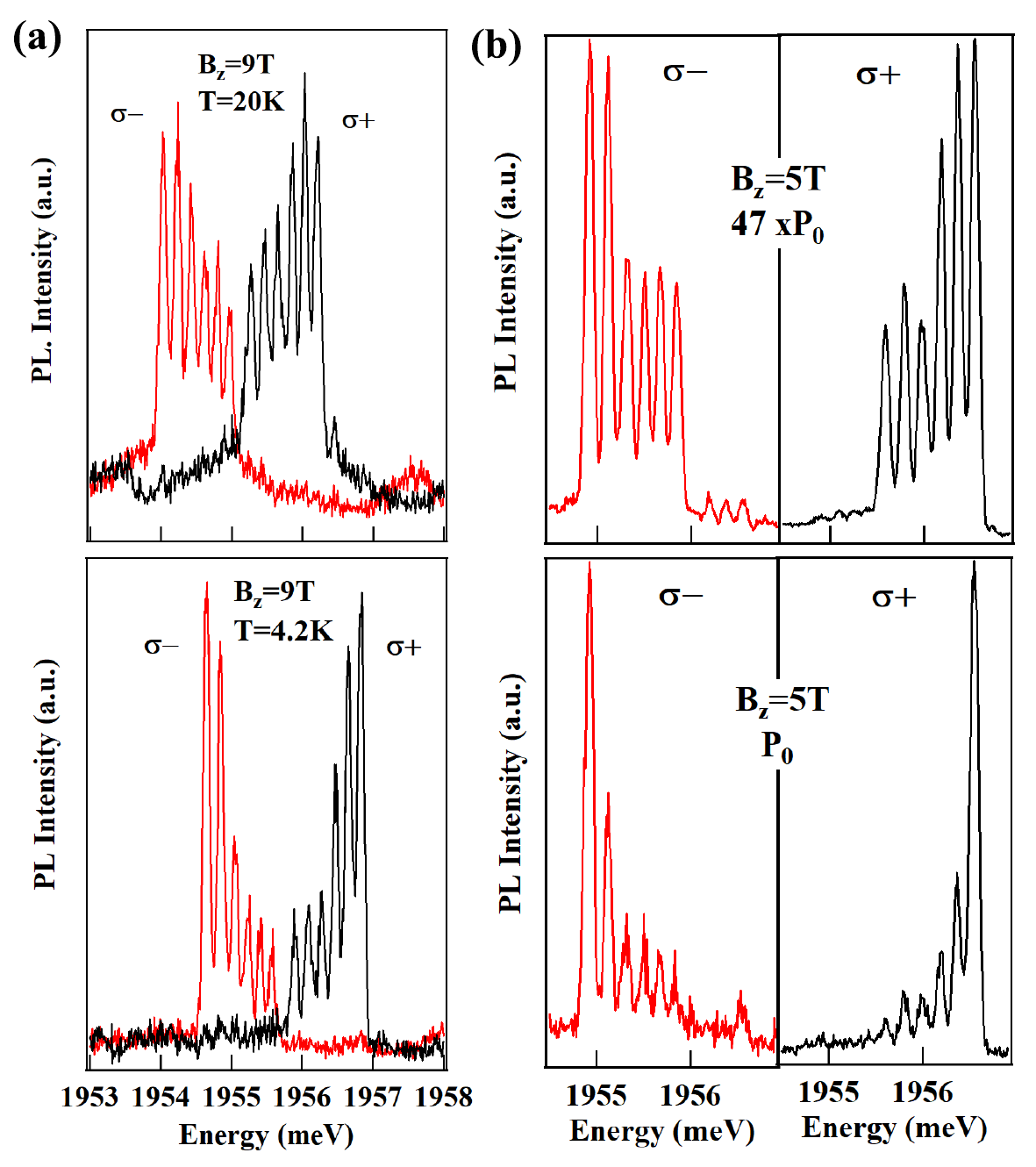}
\caption{PL intensity distribution of QDS1 under a longitudinal magnetic field. (a) Temperature dependence at a fixed excitation power under B$_z$=9T. (b) Excitation power dependence at T=4.2 K under B$_z$=5 T.}
\label{Fig8S}
\end{figure}

The distribution of intensity on the PL lines observed under magnetic field depends on the excitation intensity. This is illustrated in figure \ref{Fig8S} for QDS1 at T=4.2K under a magnetic field B$_z$=5T and for an excitation on an excited state of the dot. The thermalization on the low energy line in $\sigma-$ polarization and on the high energy line in $\sigma+$ polarization is significantly enhanced at very low non-resonant excitation intensity.

The PL intensity distribution is also significantly affected by the lattice temperature. This is illustrated in figure \ref{Fig8S} for QDS1 under B$_z$=9T, changing the temperature from 4.2 K to 20 K.

The origin of this effective spin temperature will have to be investigated. It can be controlled by interaction by non-equilibrium phonons generated during the optical excitation but can also be influenced by spin-flips among the electron-Cr$^+$ system when the QD is occupied by the positively charged exciton. The spin relaxation channels of X$^+$-Cr$^+$ will have to be studied in more details.

\section{Magneto-optics of a positively charged Mn$^{2+}$-doped QD.}

\begin{figure}[h!]
\centering
\includegraphics[width=0.8\textwidth]{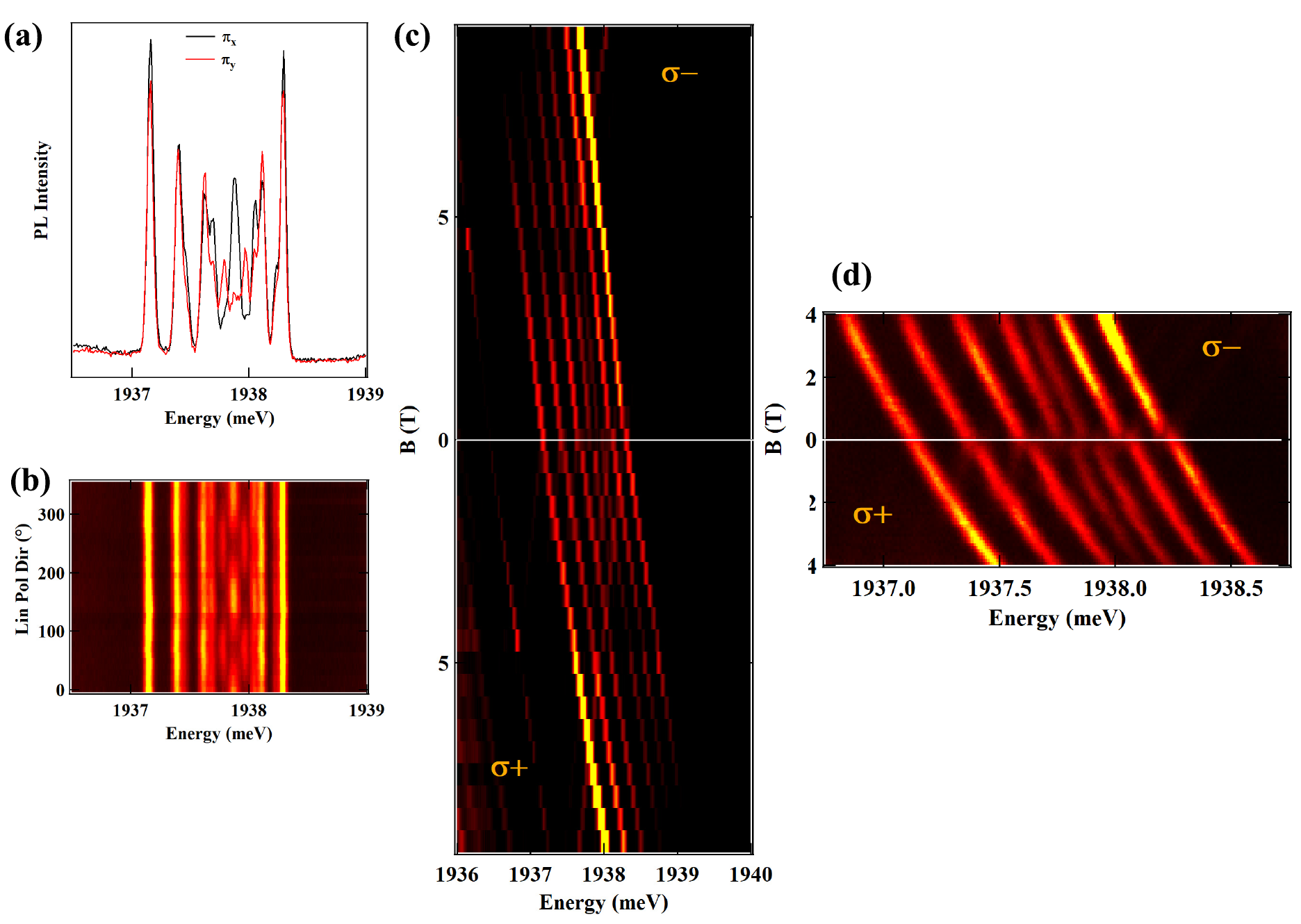}
\caption{(a) Linearly polarized PL spectra of X$^+$-Mn$^{2+}$ recorded along two orthogonal directions. (b) Linearly polarized PL intensity map. (c) Intensity map of the PL of a positively charged Mn$^{2+}$-doped QD.  (d) Detail of the PL intensity map at low magnetic field.}
\label{Fig9S}
\end{figure}

We present in Fig.~\ref{Fig9S} the main magneto-optic properties of the well known Mn$^{2+}$ system coupled with a heavy-hole in a CdTe/ZnTe QD. For a direct comparison with the data presented in the main manuscript, these magneto-optic experiments were realized on the same setup, in the same conditions and on similar CdTe/ZnTe QDs samples doped with Mn.

Positively charged Mn-doped CdTe/ZnTe QDs are characterized by (i) an anti-ferromagnetic exchange interaction between the hole and the 3$d^5$ electrons of the Mn$^{2+}$ and (ii) a X$^+$-Mn$^{2+}$ energy level structure dominated by the ferromagnetic electron-Mn$^{2+}$ coupling (I$_{eMn^{2+}}\gg$ D$_0$ and E)\cite{Lafuente2015,Lafuente2017}. The anti-ferromagnetic hole-Mn$^{2+}$ coupling appears in the intensity distribution under magnetic field with a maximum of PL on the high energy side in $\sigma-$ polarization and low energy side in $\sigma+$ polarization. This is opposite to the Cr$^+$ case studied in this article. The appearance of linear polarization at zero field in the center of the spectra is a consequence of the VBM and of the dominant exchange coupling of the isotropic electron and Mn$^{2+}$ spins within X$^+$-Mn$^{2+}$. Let us also note that because of the opposite sign of the hole-magnetic atom exchange interaction, the VBM induced gap in the PL spectra is shifted towards high energy in the case of Mn$^{2+}$ and towards low energy in the case of  Cr$^+$.


\begin{thebibliography}{}
\bibitem{note1} The intensity of the third line is weaker showing that it starts to be split.
\end{thebibliography}
\end{document}